\documentclass[twocolumn,floatfix,amsmath,amssymb,pra,showpacs,longbibliography]{revtex4-1}
\usepackage{graphicx}
\usepackage{color}
\usepackage{bm}
\usepackage{amsmath}

\newif\ifmirror
\mirrortrue
\mirrorfalse

\begin{document}

\title{Nonlinear mode competition and symmetry-protected power oscillations in topological lasers}

\author{Simon Malzard}
\affiliation{Department of Physics, Lancaster University, Lancaster, LA1 4YB, United Kingdom}

\author{Henning  Schomerus}
\affiliation{Department of Physics, Lancaster University, Lancaster, LA1 4YB, United Kingdom}

\begin{abstract}
Topological photonics started out as a pursuit to engineer systems that mimic fermionic single-particle Hamiltonians with symmetry-protected modes, whose number can only change in spectral phase transitions such as band inversions.
The paradigm of topological lasing, realized
in three recent experiments, 
offers entirely new
interpretations of these states, as they can be selectively amplified by distributed gain and loss.
A key question is whether such topological mode selection persists when one accounts for the nonlinearities that stabilize such systems at their working point.
Here we show that topological defect lasers can indeed stably operate in genuinely topological states. These comprise direct analogues of zero modes from the linear setting, as well as a novel class of states displaying symmetry-protected  power oscillations, which appear in a spectral phase transition when the gain is increased.
These effects show a remarkable practical resilience against imperfections, even if these break the underlying symmetries, and pave the way to harness the power of topological protection in
nonlinear quantum devices.
\end{abstract}
\maketitle

\section{Introduction}

Topological quantum devices aim to evoke states that display a unique response to external stimuli.
The underlying concepts were originally developed in a fermionic context, where they provide unified insights into diverse phenomena that range from the  quantum-Hall effect to the emergence of quasiparticles with unconventional statistics \cite{RevModPhys.82.3045,RevModPhys.83.1057}.
The robustness of the ensuing properties makes it desirable to replicate them in other types of systems. This can be a considerable challenge since true topological protection requires symmetries that constrain a system's behaviour. In fermionic systems the underlying symmetries originate from the  anticommutation relations and therefore are exact, even in presence of interactions \cite{Heinzner2005}. In bosonic systems, however, much looser constraints apply. The first generation of bosonic analogues of fermionic topological effects therefore required impressive feats of precision engineering to replicate the relevant single-particle physics for photons \cite{wang_observation_2009,hafezi_robust_2011,hafezi_imaging_2013,rechtsman_photonic_2013,lu_topological_2014}, cold atomic gases \cite{goldman_topological_2016}, exciton polaritons \cite{PhysRevX.5.011034,PhysRevX.5.031001,PhysRevLett.114.116401,PhysRevLett.120.097401} and sound
 \cite{Suesstrunk47,PhysRevLett.114.114301}. 

\begin{figure}[t]
\includegraphics[width=\linewidth]{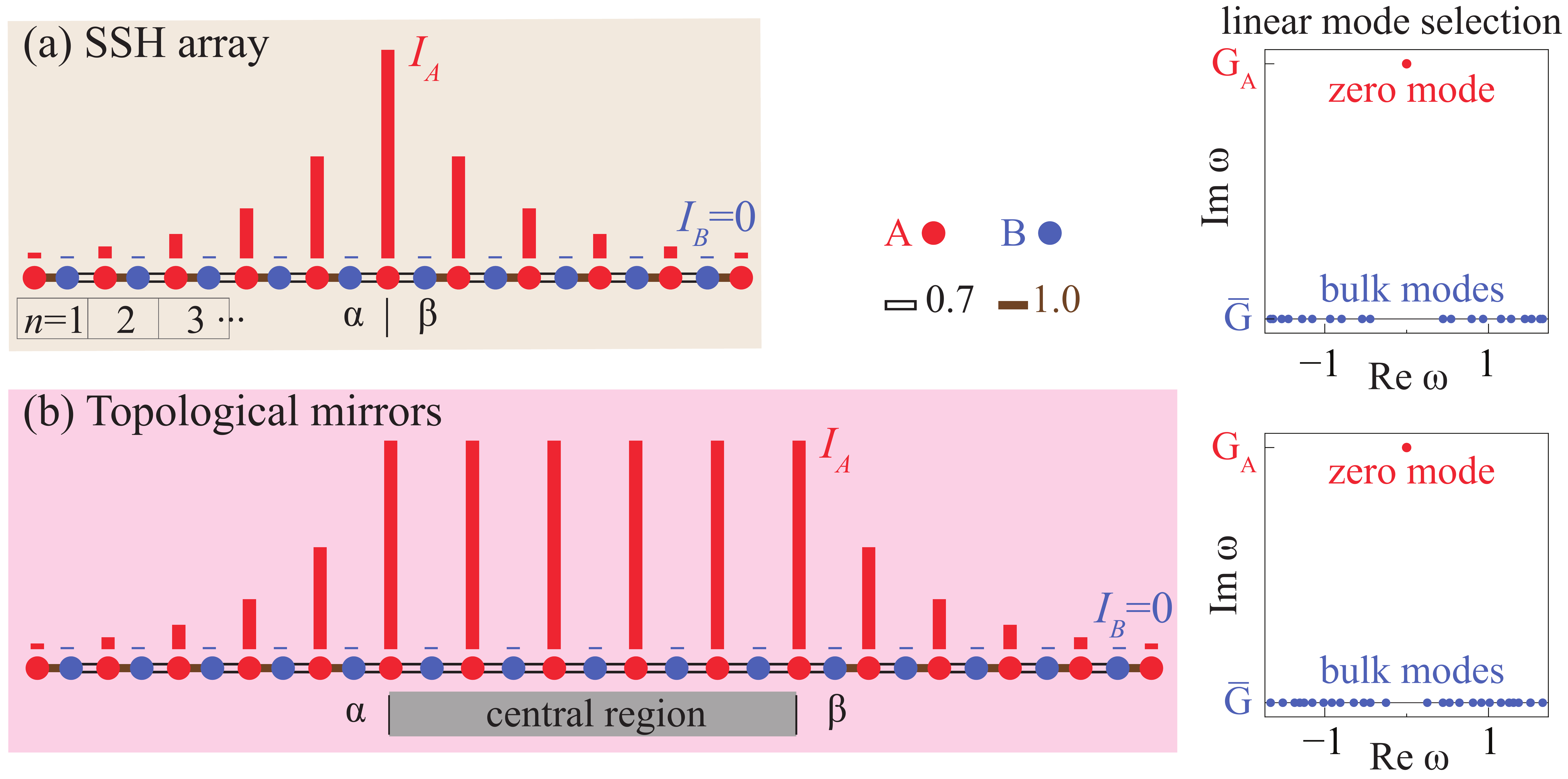}
\caption{\label{fig1}\textbf{Topological mode selection}  in laser arrays consisting of single-mode resonators grouped in dimers (enumerated by $n$). The intra-dimer couplings $\kappa$ and inter-dimer couplings $\kappa'$ are chosen to produce interfaces between regions of topologically different  band structures.  (a) In the Su-Schrieffer-Heeger (SSH) model, the alternating couplings define a phase $\alpha$ ($\kappa>\kappa'$ ) and a phase $\beta$ ($\kappa<\kappa'$).
The displayed defect state arises from two consecutive weak couplings, forming an interface between the two phases. (b)
The defect region can be extended, leading to a variant where the phases $\alpha$ and $\beta$ function as selective mirrors that confine a defect state with a larger mode volume.
In both cases, the resulting defect states have preferential weight on the A sublattice (red) and can be selected by distributed gain and loss. As illustrated in the right panels, in the linear regime the defect state acquires the effective gain $G_A$ from the $A$ sublattice, while the other modes acquire the average gain $\bar G$ in the system ($G_A=\bar G+0.1$, $\kappa,\kappa'=1,0.7$). We demonstrate that this mode selection mechanism extends to the nonlinear conditions at the working point of a laser, where it stabilizes robust zero modes and also enables alternative topological operation regimes with power oscillations.}
\end{figure}

Only in very recent times it has been realized that the liberty afforded by bosonic systems also offers opportunities for topological effects that transcend the electronic setting. Examples include squeezed light \cite{peano_topological_2016} and weakly interacting bosonic systems characterized by a Bogoliubov theory \cite{PhysRevA.88.063631,PhysRevB.87.174427,PhysRevA.91.053621,%
PhysRevB.93.020502,1367-2630-17-11-115014,PhysRevLett.115.245302,PhysRevLett.117.045302}.
The foundations for genuinely bosonic devices were laid by taking symmetries from the fermionic single-particle context and generalising them to events that change the particle number, which classically represent gain and loss \cite{PhysRevLett.102.065703,PhysRevLett.110.013903,schomerus_topologically_2013,poli_selective_2015}.

A key paradigm to test these ideas is the concept of a topological laser \cite{schomerus_topologically_2013}.
Lasing in topological defects and cavities has been realized in photonic crystals \cite{2040-8986-18-1-014005,Bahari636}, which relied on conventional mode selection, and lasing based on the analogy to topological insulators has also been put forward \cite{Hararieaar4003,Bandreseaar4005}. Separately, distributed gain and loss have been employed (beyond the conventional setting of distributed feedback lasers), e.g., in PT-symmetric lasers \cite{schomerus_quantum_2010,PhysRevLett.106.093902,Feng972,Hodaei975,Ge2016}, which exploit a spectral phase transition between conventional modes that then acquire different weights on lossy and amplified regions. Both aspects are combined  in the concept of topological mode selection
\cite{schomerus_topologically_2013,poli_selective_2015}.
This aims to utilize an additional distinctive feature of topological modes besides their pinned energy, namely that they display anomalous expectation values---typically, an unequal weight on two subspaces associated with the underlying symmetries.
This anomalous response permits to manipulate the life-time of the topological modes by
appropriate distributions of gain and loss,
which results in a topological mechanism of mode selection (see Fig.~\ref{fig1}).
Over the past months, three variants of these lasers have indeed been realised
\cite{St-Jean2017,zhao_topological_2018,PhysRevLett.120.113901}, confirming the  viability of this idea in practice.
The topological mode selection realised in these experiments directly addresses a precisely predetermined mode that exists without any further spectral phase transitions, and from the outset makes optimal use of the provided gain.

Here, we show that these systems indeed offer both genuine as well as unique topological operation conditions when one accounts for the nonlinearities that are indispensable to stabilize active systems at their working point. In particular, besides confirming the possibility of lasing in a precisely-defined nonlinear counterpart of conventional zero modes, we uncover topological phase transitions into states exhibiting topologically protected power-oscillations, not yet observed in experiments. As in conventional incarnations of topology in electronic and photonic systems, these phase transitions can be associated to structural rearrangements of spectral features, which now pertain to the excitation spectrum of the system.

These findings address key practical and conceptual challenges for the implementation and interpretation of the topological mode-selection mechanism to lasers.
On first sight it would appear that nonlinearities should degrade the effectiveness of this mechanism.
Even when starting under ideal linear conditions, the nonlinearities induce spatially varying loss and gain, which depends on the intensity profile of the mode across the system. The resulting effective gain has the potential to disfavour the topological mode, in particular when its mode volume is small.  Nonlinearities can also induce dispersive effects that explicitly break the assumed symmetries. Furthermore, even under ideal conditions where all symmetries are realized exactly, the nonlinearities can render a topological state instable, and result in spontaneous symmetry breaking and nonstationary operating regimes. 
Conceptually, we identify operation conditions that expand the practical scope of topological quantum devices and utilize nonlinear phenomena that enrich the underlying topological physics.

To establish these conclusions we evaluate the nonlinear aspects for a paradigmatic, flexible resonator arrangement and identify conditions under which topological lasing is possible. The model and its symmetries are introduced in Sect.~\ref{sec:model}. Section~\ref{sec:results1} describes the mode competition for the gain and loss distributions for which topological mode selection was originally proposed. We find that this indeed supports stationary lasing  in topological modes over wide ranges of parameter space, but also opens up a phase transition to an operation regime that exhibits topologically protected power oscillations.
Section~\ref{sec:results2} describes modified gain distributions and a setup in which the zero mode has a larger mode volume, which both offer additional means to control the operation conditions.
Section~\ref{sec:results3} considers the role of disorder and symmetry-breaking nonlinearities, which in large parts of parameter space turn out to be surprisingly tame, but also can induce phase transitions to additional operation regimes.
Paired with general considerations on the stability of nonlinear systems, these results allow us to draw  conclusions about the scope of topological effects in classical nonlinear wave dynamics, which are
described in our concluding Sect.~\ref{sec:discussion}.

\section{\label{sec:model}Nonlinear topological laser arrays}

\subsection{Modelling laser arrays with saturable gain}
The general design of the topological laser arrays studied in this work is shown in Fig.~\ref{fig1}. The arrays can be interpreted as chains of identical single-mode resonators, denoted by dots, which are coupled evanescently to their nearest neighbors. Given this structure of the coupling it is convenient to divide the system into two alternating sublattices A and B, and group neighbouring pairs of A and B sites into dimers. Denoting the corresponding wave amplitudes on the $n$th dimer
as  $A_n$ and $B_n$, their dynamical evolution is then governed by the coupled-mode equations
\begin{subequations}
\begin{align}
i\frac{dA_n}{dt}=[\omega_{A,n}+V_{A,n}(|A_n|^2) ]A_n+ \kappa_{n}B_n+\kappa'_n B_{n-1},\\
i\frac{dB_n}{dt}=[\omega_{B,n}+V_{B,n}(|B_n|^2) ]B_n+ \kappa_{n}A_n+\kappa'_{n+1} A_{n+1},
\end{align}
\label{eq:model}%
\end{subequations}
where $\omega_{s,n}$ ($s=A,B$) are the bare resonance frequencies of the isolated resonators, $\kappa_n$ is the intra-dimer coupling between the A and B site in  the $n$th dimer, and $\kappa'_n$ is the inter-dimer coupling between the B site in the $(n-1)$st dimer and the A site in the $n$th dimer
\footnote{These couplings
can always be made positive by a suitable $\mathbb{Z}_2$ gauge transformation.}.
The effective complex potentials \cite{PhysRevA.72.013803}
\begin{subequations}
\begin{align}
V_{A,n}(|A_n|^2)=(i+\alpha_A)\left(\frac{g_A}{1+S_A|A_n|^2}-\gamma_A\right),\\
V_{B,n}(|B_n|^2)=(i+\alpha_B)\left(\frac{g_B}{1+S_B|B_n|^2}-\gamma_B\right)
\end{align}
\label{eq:potentials}%
\end{subequations}
model nonlinear saturable gain of strength $g_s$ and background loss $\gamma_s$, where the real constants $S_s$ and $\alpha_s$ are the self-saturation coefficient and the linewidth-enhancement (or anti-guiding) factor, respectively.

\subsection{\label{sec:linselection}Topological features and mode selection in the linear regime}
For identical passive resonators  $\omega_{s,n}\equiv \omega_{AB}$ with vanishing gain and loss ($g_s=\gamma_s=0$) and an alternating coupling sequence $\kappa_n\equiv \kappa$, $\kappa_n'\equiv \kappa'$, the array is an  incarnation of the celebrated Su-Schrieffer-Heeger (SSH) model \cite{PhysRevLett.42.1698,ryu_topological_2002}. This model displays a symmetric band structure with a gap of size $\Delta =2|\kappa-\kappa'|$ around the central frequency $\omega_{AB}$, which induces two topological phases  $\alpha$ (where $\kappa>\kappa'$) and $\beta$ (where $\kappa<\kappa'$).
At a physical interface between these phases one encounters a localized defect mode  [see Fig. \ref{fig1}(a)] whose frequency $\Omega_0=\omega_{AB}$ is pinned to the centre of the gap. Due to its topological origin this mode persists for more complicated interface configurations, which we will exploit to change its mode volume as shown in  Fig. \ref{fig1}(b). There, the terminating dimer chains operate as topological mirrors while the defect mode extends uniformly over the central part of the system.

Exact zero-mode quantization  requires that the system is terminated on a fixed sublattice (here the A sublattice), which then contains one more site than the other sublattice (here B), so that overall the system supports an odd number of modes.
An intimately related principal feature of this topological mode is that it only occupies the majority sublattice A (so that $\mathbf{B}=0$), in contrast to all other states in the system which have equal weight on both sublattices ($|\mathbf{A}|=|\mathbf{B}|$; see the Appendix for a short proof of these spatial features). The defect mode can therefore be addressed by sublattice-dependent gain and loss, which results in a simple and robust mode-selection mechanism that employs the topological origin of the mode.
Assuming that the gain and loss are linear ($S_s=0$) and do not break the symmetry of the frequency spectrum ($\alpha_s=0$),
the topological mode then acquires the effective gain $G_A=g_A-\gamma_A$  on the A sublattice, while all bulk modes acquire the average effective gain $\bar G=(g_A+g_B-\gamma_A-\gamma_B)/2$ (see the right panels in Fig.~\ref{fig1}). The topological protection persists because the effective non-hermitian Hamiltonian exhibits a non-hermitian charge-conjugation symmetry $(H-\omega_{AB})^*=-\sigma_z (H-\omega_{AB})\sigma_z$ (with the Pauli matrix $\sigma_z$ operating in sublattice space), which stabilizes any complex eigenvalues $\Omega_n$ positioned on the axis $\mathrm{Re}\,\Omega_n=\omega_{AB}$  \cite{Pikulin2012a,PhysRevB.87.235421,schomerus_topologically_2013,poli_selective_2015,malzard_topologically_2015,sanjose2016}.

\begin{figure*}[bt]
\includegraphics[width=\linewidth]{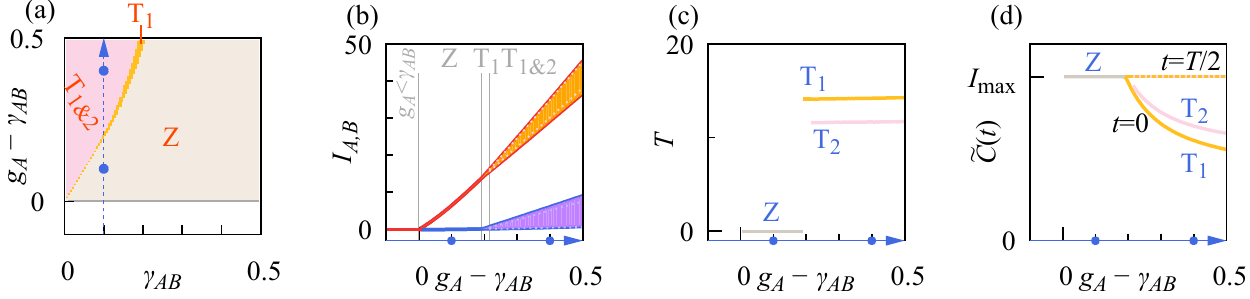}
\caption{\label{fig2a} \textbf{Topological lasing regimes} for the SSH array of Fig.~\ref{fig1}(a) pumped on the A sublattice (finite gain $g_A$ at fixed $g_B=0$, with amplitudes scaled such that $S_A=S_B=1$) under conditions that preserve the symmetries in the linear case ($\omega_{s,n}=\omega_{AB}$, $\alpha_s=0$), demonstrating operation in topological states over the whole parameter range.
(a) Phase diagram of stable quasistationary operation regimes depending on the gain $g_A$ and background losses $\gamma_A=\gamma_B\equiv\gamma_{AB}$, where lasing requires  $g_A>\gamma_{AB}$.
Over the whole gray region labelled Z, the system establishes stationary lasing in a topological zero mode. In the orange region, this is replaced by operation in a twisted topological mode T$_1$ displaying power oscillations. In the pink region an additional twisted state T$_2$  exists, whose selection then depends on the initial conditions.
The remaining
panels analyze the lasing characteristics for varying gain $g_A$ along the line $\gamma_{AB}=0.1$ (blue arrow in the phase diagram). (b) Sublattice-resolved intensities $I_A$ (red) and $I_B$ (blue), including shaded intensity ranges for the power oscillations of T$_1$ and dashed lines indicating the corresponding ranges for T$_2$. (c) Amplitude oscillation period $T$ (equalling twice the period of power oscillations for twisted states, see Fig.~\ref{fig2b}).  (d) Correlation function $\tilde C(t)$ at $t=0, T/2$, where $\tilde C(T/2)=I_{\rm max}$ reveals the topological nature of the states (see text). As illustrated for the examples in Fig.~\ref{fig2b}, all states inherit the intensity profile of the linear defect mode from Fig.~\ref{fig1}(a).}
\end{figure*}

\subsection{Nonlinear extension of charge-conjugation symmetry}

While this linear mechanism describes an initial competitive advantage of the defect mode if $G_A>\bar G$, and allows it to dynamically switch on if $G_A>0$ and the intensity is still small, this does not describe the quasi-stationary operation regime where the medium saturates in response to a much larger intensity. This saturation is critical for the stabilization of any laser at its working point, where the medium provides just as much energy as the lasing mode loses through radiative and absorptive processes.

The nonlinear modification \eqref{eq:potentials} of the model includes the saturation dynamics and allows us to address the  key challenges for topological lasing described in the introduction.
(i) The self-saturation quantified by $S_s$ makes the effective gain or loss nonuniform across the whole system, while at the same time favouring modes with a large mode volume.
(ii) The linewidth-enhancement factor $\alpha_s$ induces symmetry-breaking terms in the gain and loss, which we can compare with linear symmetry-breaking via disorder in the bare resonance frequencies $\omega_{s,n}$.
(iii) The system admits a dynamical counterpart of the non-hermitian charge-conjugation symmetry that extends to the nonlinear case \cite{2017arXiv170506895M},  described next, which will allow us to identify a range of topological and nontopological operation regimes.

This dynamical counterpart of the non-hermitian charge-conjugation symmetry can be phrased in terms of the following general property of the coupled-mode equations \eqref{eq:model}.
For any set of parameters $(\omega_{s,n},\kappa_n,\kappa_n',g_s,\gamma_s,\alpha_s,S_s)$ and an arbitrarily chosen reference frequency $\omega_{AB}$, any solution $\Psi(t)=\left(\begin{array}{c}\mathbf{A}(t)\\\mathbf{B}(t)\end{array}\right)$ can be mapped onto a solution
\begin{equation}
\label{eq:tildepsi}
\tilde\Psi(t)=\exp(-2i\omega_{AB} t) \left(\begin{array}{c}\mathbf{A}^*(t)\\ -\mathbf{B}^*(t)\end{array}\right)
\end{equation}
of the model with parameters replaced by $(\tilde\omega_{s,n}= 2\omega_{AB}-\omega_{s,n},\kappa_n,\kappa_n',g_s,\gamma_s,-\alpha_s,S_s)$.
This property turns into a dynamical symmetry if $\omega_{s,n}=\omega_{AB}$, $\alpha_s=0$ (hence, the same conditions as observed for the non-hermitian charge-conjugation symmetry in the linear case).

To define the resulting operation regimes we henceforth set the reference frequency to $\omega_{AB}\equiv 0$, which can always be achieved through a gauge transformation $\Psi(t)\to \Psi(t)\exp(-i\omega_{AB}t)$.
We then can distinguish
self-symmetric stationary states $\Psi(t)=\tilde \Psi(t)=\mathrm{const}(t)$, which we interpret as nonlinear topological zero modes [Z], time-dependent versions of such self-symmetric states [S], as  well as the notable class of twisted modes where $\Psi(t+T/2)=\tilde \Psi(t)$ [T]. The twisted modes are automatically periodic,  $\Psi(t+T)=\Psi(t)$, and hence lead to stable power oscillations of period $T/2$.  Finally, we can also encounter stationary and time-dependent lasing modes that spontaneously break the dynamical symmetry, which automatically occur in pairs $\Psi(t)$,  $\tilde \Psi(t)$ [P].

To discriminate between these types of modes we utilize the correlation functions
\begin{subequations}%
\begin{align}
& C(t)=|\langle \Psi(t_{\rm max})|\Psi(t_{\rm max}+t)\rangle |,\\
&\tilde C(t)=|\langle \tilde \Psi(t_{\rm max})|\Psi(t_{\rm max}+t)\rangle |.
\end{align}%
\end{subequations}%
For periodic modes $t_{\rm max}$ will be chosen such that $C(0)=I_{\rm max}$ coincides with the intensity maximum over a period. For stationary modes $t_{\rm max}$ is arbitrary and $I_{\rm max}$ is to be interpreted as the stationary intensity.
Self-symmetric modes are  characterized by coinciding correlation functions, $\tilde C(t)=C(t)$. For twisted modes the correlation functions alternate with an offset $T/2$, hence $\tilde C(t)=C(t+T/2)$ and in particular $\tilde C(T/2)=I_{\rm max}$. For symmetry-breaking modes, the two correlation functions do not bear any simple relation but are constrained by $\tilde C(t)<I_{\rm max}$ for all $t$.

\section{\label{sec:results1}Ideal topological lasing}
\subsection{Operation regimes}

We first consider lasing under the ideal conditions under which topological mode selection was originally conceived. 
This requires laser arrays with exact  non-hermitian charge-conjugation symmetry ($\omega_{s,n}\equiv \omega_{AB}= 0$, $\alpha_s=0$) and gain confined to the A sublattice ($g_A$ finite and variable by the pumping, while $g_B=0$). We consider resonator-independent self-saturation coefficients and scale the amplitudes $\textbf{A}$ and $\textbf{B}$ such that $S_A=S_B=1$. The operation of the system then depends on the balance between the gain and the  linear background losses, which we here assume to be resonator-independent and denote as $\gamma_A=\gamma_B\equiv\gamma_{AB}$.

Figure \ref{fig2a} provides an overview of the resulting operation regimes. Over a large part of the parameter space, the laser operates in a stable zero mode, which is quickly approached over time irrespective of initial conditions. In the phase diagram (panel a), this region is indicated by the label Z. When the gain/loss ratio is increased the zero mode becomes unstable and is replaced by a twisted mode T$_1$, which results in lasing with power oscillations.  As we show in Sec.~\ref{sec:phasetransition}, this change comes about in a topological phase transition. Upon a small further increase of the gain/loss ratio the mode T$_1$ starts to compete with a second twisted mode T$_2$. Both modes sustain stable lasing with power oscillations of different amplitude and period, where the choice of mode depends on the initial conditions.

Panels (b-c) in Fig. \ref{fig2a}  examine the key characteristics of the lasing modes as one varies the gain for fixed background losses $\gamma_{AB}=0.1$, which covers all described regimes.
The gain-dependence of the sublattice-resolved intensities $I_A=|\mathbf{A}|^2$ and $I_B=|\mathbf{B}|^2$
can be interpreted as light-light curves. $I_A$ displays a characteristic kink as one crosses the laser threshold $g_A=\gamma_{AB}$ and enters stationary operation in the zero mode Z, while $I_B$ initially remains negligible. Upon increasing the gain, the stationary operation regime is replaced by lasing in the twisted mode T$_1$, which from its onset displays a finite period $T$ while the amplitude of its power oscillations (of period $T/2$) increase smoothly (see the shaded intensity ranges). The second twisted mode sets in with a slightly smaller period, but covers very similar intensity ranges (indicated by the dashed white lines).

Panel (d)  in Fig. \ref{fig2a}  verifies the symmetry-protected nature of these states throughout the whole range of gain. For the self-symmetric zero mode Z, this is evidenced by its characteristic property $\tilde C(0)=I_{\rm max}$. The twisted modes are not self-symmetric, $\tilde C(0)<I_{\rm max}$, but display their hallmark property $\tilde C(T/2)=I_{\rm max}$.

\begin{figure}[t]
\includegraphics[width=\linewidth]{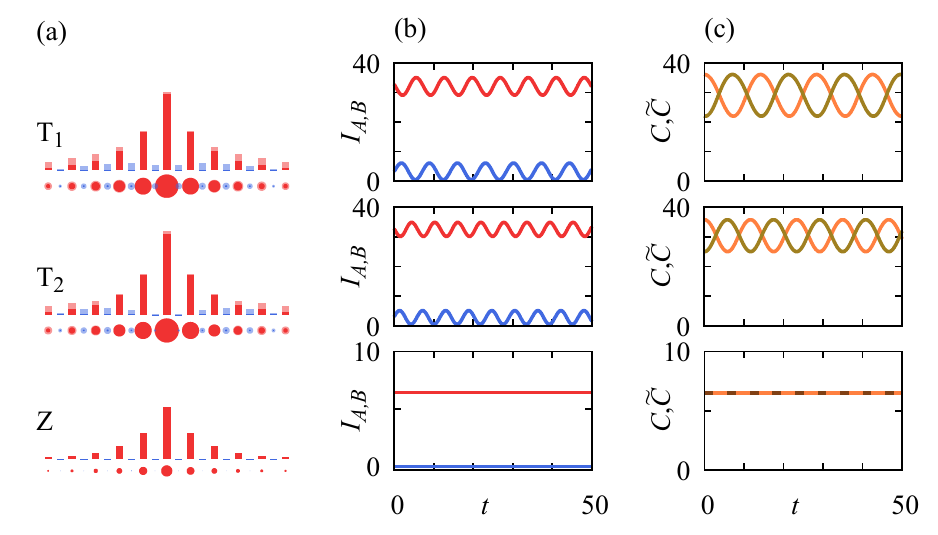}
\caption{\label{fig2b} \textbf{Topological wave features} of representative lasing states at parameters indicated by blue dots in Fig.~\ref{fig2a} ($g_A=0.2$ for $Z$, $g_A=0.5$ for T$_1$ and T$_2$).
(a) Intensity distributions over the array, shown both as spikes and as disks with area proportional to intensity, substantiating the relation of these stabilized states to the linear defect state from Fig.~\ref{fig1}(a).
(b,c) Time-dependence of the sublattice-resolved intensities  $I_A(t)$ and $I_B(t)$ (red and blue) and of the correlation functions $C(t)$, $\tilde C(t)$ (orange and brown).
The alternating correlations $\tilde C(t+T/2)=C(t)$ verify the twisted nature of the states T$_1$ and T$_2$, while $C(t)=\tilde C(t)=\mathrm{const}$ verifies that the state  Z is a topological zero mode.}
\end{figure}

These features are further corroborated by the examples of modes shown in Fig.~\ref{fig2b}. As shown in panel (a), all modes clearly inherit their profile from the linear defect state of Fig.~\ref{fig1}(a). For the stationary mode, the intensity $I_B$ on the B sublattice is very small and practically negligible.
For the twisted modes, $I_B$ is of the order of the associated power oscillations. Note that the intensities on both sublattices oscillate out of phase (see panel b), and so do the correlation functions $C(t)$ and $\tilde C(t)$ (panel c), as required by their twisted nature. Furthermore, the period $T/2$ of the power oscillations in $I_{A,B}(t)$ is indeed half of that of the amplitude oscillations exhibited by the amplitude correlation functions.

\begin{figure}[t]
\includegraphics[width=\linewidth]{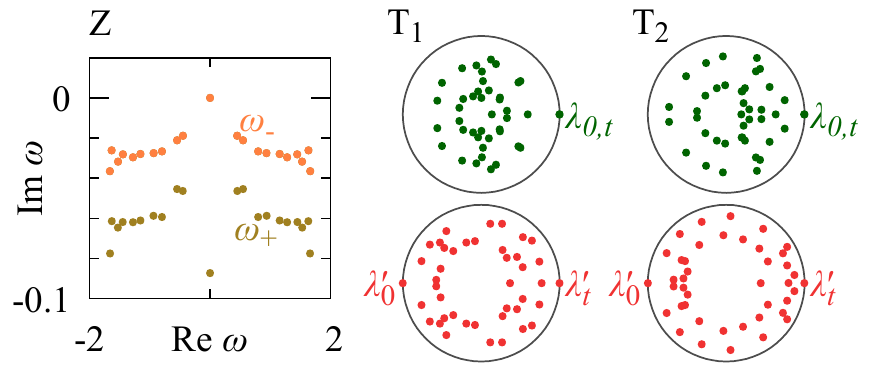}
\caption{\label{fig2c} \textbf{Stability excitation spectra} of the representative states illustrated in Fig.~\ref{fig2b}.
For the stationary state Z this represents the Bogoliubov spectrum $\omega$, which separates into excitations $\omega_\pm$ that preserve or break the symmetry. This separation further verifies its zero-mode character (see text), while $\mathrm{Im}\,\omega< 0$ [apart from the U(1) Goldstone mode at $\omega=0$] affirms that the state is stable.
For the periodically oscillating states T$_1$ and T$_2$, this represents the Bogoliubov-Floquet stability spectrum $\lambda$ (top, green) and the spectrum $\lambda'$ of the half-step propagator (bottom, red). Both spectra are confined by the unit circle in the complex plane, demonstrating that these states are stable. The symmetry-protected excitations pinned to $\lambda'=\pm 1$ further verify the twisted nature of these states.}
\end{figure}

\subsection{\label{sec:phasetransition}Topological excitations and phase transition}

\begin{figure}[t]
\includegraphics[width=\linewidth]{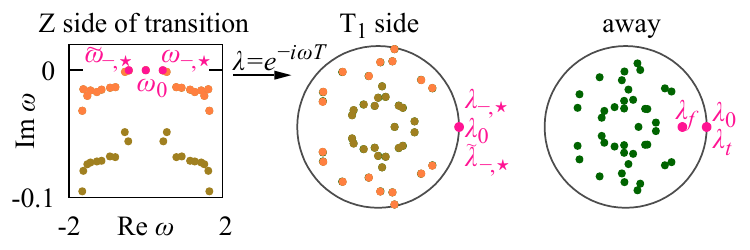}
\caption{\label{fig2d} \textbf{Topological phase transition} between the zero-mode regime Z and the twisted mode T$_1$, at $g_A=0.291$ along the line $\gamma_{AB}=0.1$ (see Fig.~\ref{fig2a}). At the transition two Bogoliubov excitations $\omega_{-,\star}=2\pi/T$ and ${\tilde\omega}_{-,\star}=-\omega_{-,\star}$ are marginally stable, where $T$ is the period of the emerging twisted mode T$_1$. Along with the U(1) Goldstone mode, they all map onto  Floquet-Bogoliubov excitations $\lambda=1$ for this emerging mode. Away from the transition, these excitations split into two degenerate excitations $\lambda_0=\lambda_t=1$ associated with the U(1) and time translation freedoms, and a decaying excitation $\lambda_f$ related to the amplitude stabilization of the power oscillations. (Note that at the transition another pair of excitations is almost unstable, which will give rise to the twisted mode T$_2$.) }
\end{figure}

The fact that we find a well-defined set of lasing modes professes that the described topological states are very stable, at least as long as one stays away from the phase transition between the operation regimes. This can be ascertained by a linear stability analysis, which results in a complex Bogoliubov excitation spectrum $\omega_n$. For a time-periodic mode, a similar analysis can be carried out based on a Bogoliubov-Floquet propagator $F$ of excitations over an oscillation period, whose eigenvalues are written as $\lambda_n=\exp(-i\omega_n T)$. These excitation spectra  reveal intriguing topological features, which are illustrated in Fig.~\ref{fig2c} for the three example modes of Fig.~\ref{fig2b}, and in Fig.~\ref{fig2d} at the phase transition between the operation regimes. We here describe the resulting phenomenology, while the technical details are recapitulated in the Appendix. The key feature of this discussion is the concept topological excitations that are pinned to symmetry protected positions, in analogy to Majorana zero modes in fermionic systems with charge-conjugation symmetry \cite{RevModPhys.82.3045,RevModPhys.83.1057} and zero modes in periodically driven systems \cite{PhysRevB.82.115120,Kitagawa2012}.

As the coupled-mode equations describe the dynamics of a complex wave, the stability spectra contain twice as many excitations as there are states in the linear system. These excitations are constrained by an independent spectral symmetry, forcing them to obey $\mathrm{Re}\,\omega_n=0$ or to occur in pairs $\omega_n$, $\tilde\omega_n=-\omega_n^*$ (equivalently, the Bogoliubov-Floquet eigenvalues $\lambda_n$ are either real or form complex-conjugated pairs $\tilde \lambda_n=\lambda_n^*$).
A state is stable if all physical excitations decay, $\mathrm{Im}\,\omega_n<0$ (hence $|\lambda_n|<0$). An exception is the U(1) Goldstone mode pinned at $\omega_0=0$ ($\lambda_0=1$), which arises from the arbitrary choice of the global phase of the wavefunction $\Psi$. This phase can diffuse due to quantum noise, which results in the finite linewidth of the emitted laser light. Furthermore, in the Floquet case an additional pinned eigenvalue $\lambda_t=1$ arises from the arbitrary choice of the reference time $t_0$ for any solution $\Psi(t+t_0)$. Let us now examine how this general picture is modified by topological excitations.

As illustrated in Fig.\ \ref{fig2c}, the symmetries of the topological states allow us to systematically deconstruct their excitation spectrum. For the stationary zero mode Z, we can distinguish excitations that preserve the self-symmetry, denoted as $\omega_{+,n}$, from
excitations that break the self-symmetry, denoted as $\omega_{-,n}$. The latter contain the Goldstone mode $\omega_{-,0}=0$, and in our setting describe the more slowly decaying excitations.
Notably, in the considered system both sets of spectra contain an odd number of excitations (equalling the number of resonators in the laser array).

As further illustrated in the figure, for the twisted states T$_1$ and  T$_2$ we can relate the eigenvalues $\lambda_n={\lambda_n'}^2$ to the eigenvalues of a twisted half-step propagator $F'$ that characterizes propagation of excitations over half a period $T/2$ (see the Appendix for the exact definition of this propagator). This reduced spectrum contains a mode pinned at ${\lambda}_t'=1$, which arises from time-translation invariance, and a mode
pinned to $\lambda_0'=-1$, which originates from the U(1) Goldstone mode. This configuration of excitations for propagation over half a period constitutes a distinctive topological signature of the twisted modes.

The described features become are further illuminated when one inspects the phase transition between the zero-mode regime and the twisted state T$_1$.
In the general setting of nonlinear optical systems \cite{PhysRevE.56.6524,doi:10.1063/1.1480915}, this transition corresponds to a Hopf bifurcation, which here however occurs in a symmetry-constrained setting.
Figure~\ref{fig2d}(a) shows the Bogoliubov spectrum at the transition, where a pair of symmetry-breaking excitations with $\tilde \omega_{-,\star}=-\omega_{-,\star}$ crosses the real axis and thereby destabilizes the zero mode. This pair of excitations
combines to display the oscillatory time dependence of the emerging twisted state T$_1$, whose initial oscillation frequency is given by $2\pi/T=|\omega_{-,\star}|$. Different combinations of these two excitations amount to a time translation of these resulting oscillations. Notably, at the transition the Bogoliubov-Floquet spectrum of this emergent state is given by $\lambda_n=\exp(-i\omega_n T)$, as is illustrated in Fig.~\ref{fig2d}(b).

Note that upon this mapping the destabilizing excitations $\tilde \omega_{-,\star}=-\omega_{-,\star}$ map to $\lambda_{-,\star}=\tilde \lambda_{-,\star}=1$. For the twisted mode, they therefore constitute two excitations that right at the transition are both degenerate with the U(1) Goldstone mode.
Departing from the transition into the twisted-state regime [Fig.~\ref{fig2d}(c)], these excitations split into two separate real eigenvalues $\lambda_t$ and $\lambda_f$. Of these, $\lambda_t$  describes the time-translation freedom and therefore remains degenerate with the U(1) Goldstone mode. The eigenvalue $\lambda_f$, on the other hand, is associated with perturbations of the finite amplitude of the power oscillations. These perturbations decay due to the nonlinear feedback, so that $|\lambda_f|<1$, guaranteeing that  the oscillations are stable. This mechanism gives rise to the aforementioned topological excitations $\lambda_0'=-1$, $\lambda_t'=1$ in the half-step propagator, which remain a robust signature of the twisted state even when one moves far away from the transition, as we already have seen in the examples of Fig.~\ref{fig2c}.

\begin{figure}[t]
\includegraphics[width=\linewidth]{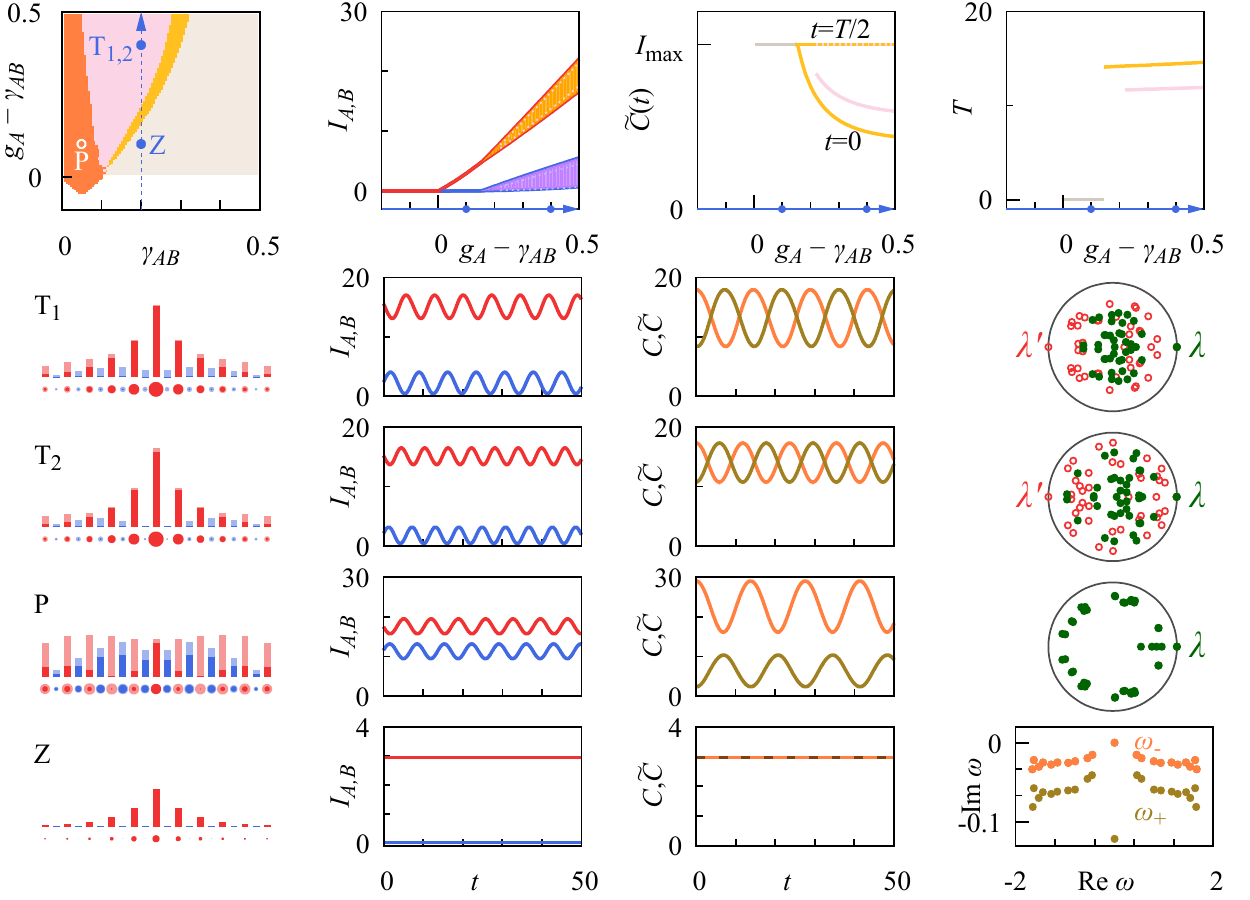}
\caption{\label{fig3}
\textbf{Role of reduced gain imbalance}, obtained under the same conditions as in Figs.~\ref{fig2a}-\ref{fig2c} (see also Fig.~\ref{fig:2summary}), but with finite gain $g_B=0.1$ on the $B$ sublattice. For $\gamma_{AB}<g_B$ the parameter space now also contains a region (dark orange) supporting additional pairs of symmetry-breaking modes P. As illustrated for the marked example, these modes have substantial weight on the B sublattice, while their independent correlation functions $C(t)$ and $\tilde C(t)$ show that they spontaneously break the symmetry. For such modes the Bogoliubov-Floquet spectrum contains many eigenvalues close to the unit circle, indicating their high sensitivity under parameter changes. As shown in the top panels for the cross-section now placed at $\gamma_{AB}=0.2$, the remaining parameter space supports the same robust topological lasing modes as observed for $g_B=0$ (twisted modes T$_1$ and  T$_2$ and stationary topological modes Z, as illustrated by the marked examples).
}
\end{figure}

\begin{figure}[t]
\includegraphics[width=\linewidth]{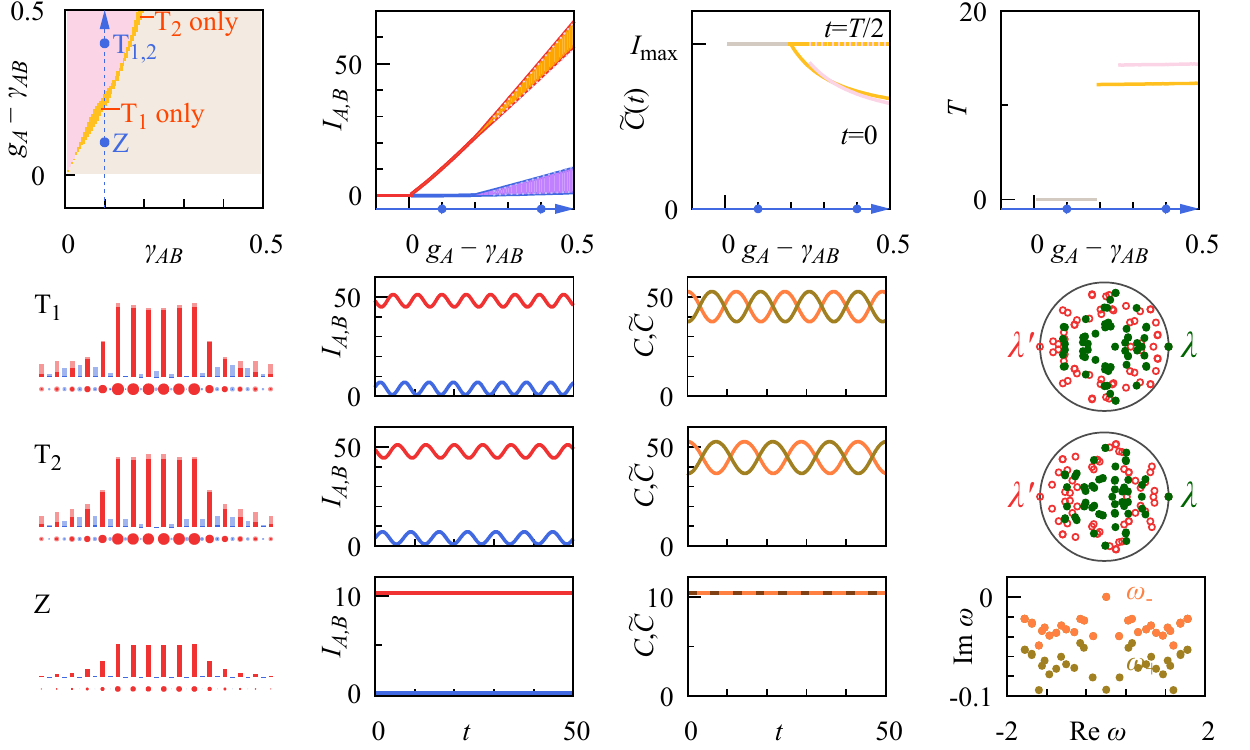}
\caption{\label{figmirror1}
\textbf{Role of increased mode volume}, obtained for the laser array with topological mirrors illustrated in Fig.~\ref{fig1}(b). Here we consider ideal lasing conditions with variable gain $g_A$ and background loss $\gamma_{A}=\gamma_{B}\equiv \gamma_{AB}$, at vanishing gain $g_B=0$ on the B sublattice.
The representation of the data is the same as in Fig.~\ref{fig3}. The resulting operation regimes closely resemble those of the SSH laser array under corresponding conditions (cf.~Figs.~\ref{fig2a}-\ref{fig2c}, summarized in Fig.~\ref{fig:2summary}), with a phase of stationary zero-mode lasing supplemented by phases with one or two twisted modes displaying power oscillations. The intensities of these modes have increased, which reflects their larger mode volume, as illustrated in more detail for the three examples marked $Z$, T$_1$ and  T$_2$.
}
\end{figure}

\begin{figure}[t]
\includegraphics[width=\linewidth]{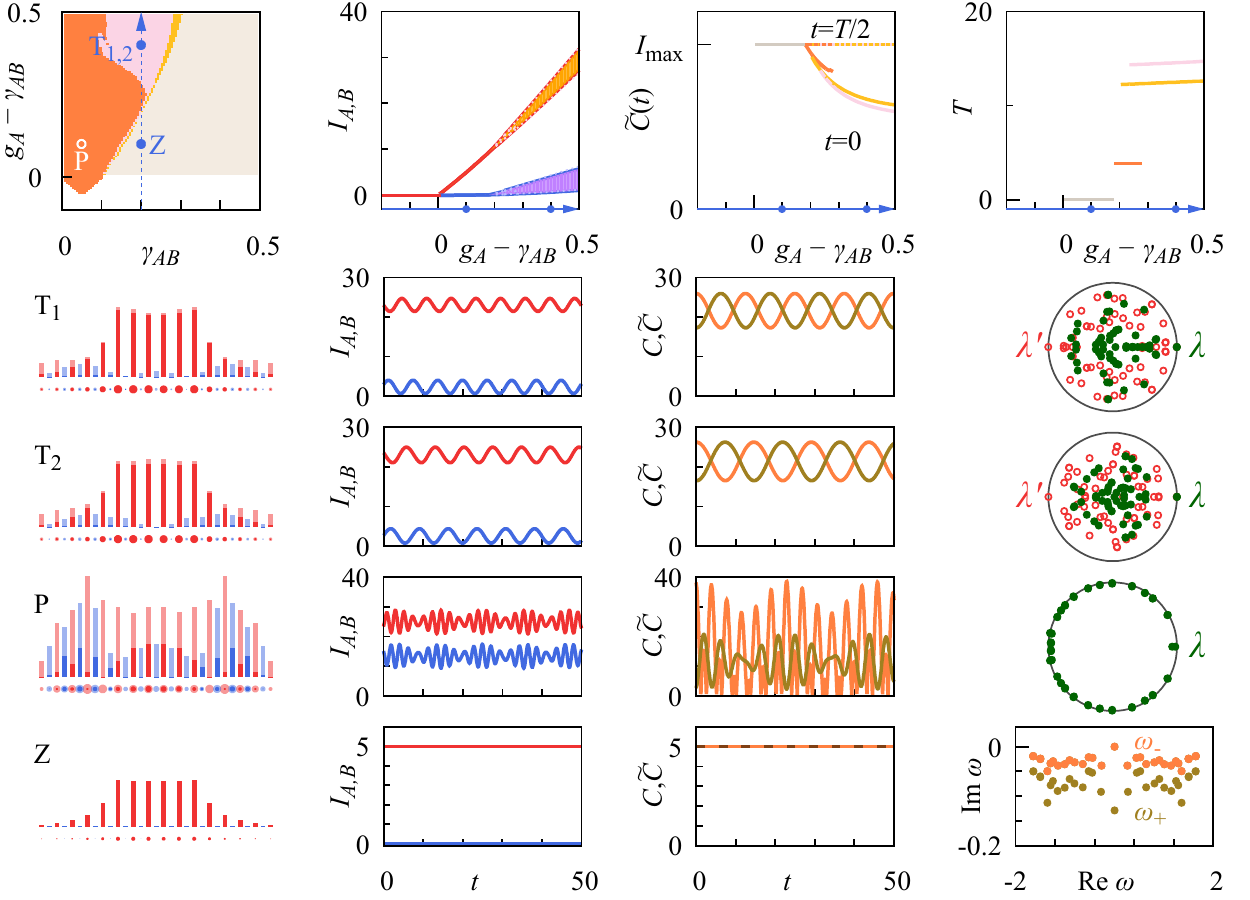}
\caption{\label{figmirror2}
\textbf{Interplay of mode volume and gain imbalance}.
Same as Fig.~\ref{figmirror1}, but for finite gain $g_B=0.1$ on the B sublattice, and the cross-section through parameter space shifted to $\gamma_{AB}=0.2$.
Compared to the corresponding conditions in the SSH laser array (Fig.~\ref{fig3}), a larger range of parameters now supports a multitude of additional states. At the representative point marked P, this includes a pair of symmetry-breaking oscillating states, whose power oscillations are modulated. The features of these symmetry-breaking  states are not very robust, as indicated by their Bogoliubov-Floquet stability spectra, which display many slowly decaying excitations. These modifications are restricted to the range of parameters that previously displayed the twisted states T$_1$ and T$_2$ (now only seen for large enough gain), but does not affect the operation in the zero-mode $Z$.
Along the cross-section $\gamma_{AB}=0.2$, we enter only briefly enter this modified regime, in a region where there is only one extra, twisted, state, which destabilizes the zero mode.}
\end{figure}

\section{\label{sec:results2}Modified operation conditions}

To verify the versatility and resilience of the laser array
we consider two ways to modify the mode competition between the different states in the system. To facilitate the comparison with ideal conditions,
Fig.~\ref{fig:2summary} in the Appendix provides a condensed summary of
Figs.~\ref{fig2a}, \ref{fig2b} and \ref{fig2c}.

\subsection{Modified gain distribution}
Figure \ref{fig3} examines the role of the gain distribution via the  addition of finite gain $g_B=0.1$ on the B sublattice, which amounts to a reduction of the gain imbalance.
In the linear model, the additional gain does not affect the defect state, which sees the effective gain $G_A$, but increases the effective gain $\bar G$ of all the other states in the system (see Fig.~\ref{fig1}). In the nonlinear model, the additional gain modifies the operation regimes in parts of the region $\gamma_{AB}<g_B$, where the losses are not strong enough to suppress modes with substantial weight on the B sublattice. Besides additional twisted modes, this region then become populated by oscillating pairs of symmetry-breaking modes P. As shown for an example in the figure,
these modes extend over the whole system and display substantial weight on both sublattices. The Bogoliubov-Floquet spectrum of any two partner modes are identical, but they cannot be further deconstructed as for the topological states. The position of the eigenvalues close to the unit circle reflects a reduced robustness of these symmetry-breaking modes against parameter variations.

In the remainder of parameter space we encounter the same topological operation regimes as in the ideal case, with the boundary between zero modes and twisted modes now shifted to larger losses. The modes themselves display the same features as before, as  illustrated for variable gain $g_A$ along the line $\gamma_{AB}=0.2$.
The threshold to stationary lasing again gives rise to a marked increase of intensity on the A sublattice, while the power oscillations of the twisted states at larger gain display very similar periods and relative amplitudes as before.
The three marked examples verify that these topological modes still inherit their mode profile from the linear defect state, and display the required topological correlations and excitations that can only change in phase transitions.

\subsection{Modified mode volume}
Figures \ref{figmirror1} and \ref{figmirror2} examine the modified set-up of Fig.~\ref{fig1}(b), where the defect region is extended. In the linear system, the terminating regions act as selective mirrors for a zero mode with an increased mode volume, which remains confined to the A sublattice. Moreover, because of its increased length the system also supports a larger number of extended states that compete for the gain. In Fig.~\ref{figmirror1} the gain on the B sublattice is set to $g_B=0$, while in Fig.~\ref{figmirror2} we have $g_B=0.1$.

In the ideal case $g_B=0$ (Fig.~\ref{figmirror1}), the resulting operation regimes closely resemble those of ideal lasing in the SSH laser array (Figs.~\ref{fig2a}-\ref{fig2c}). The parameter space is divided into a region with a topological zero mode $Z$ and regions with one or two twisted modes T$_1$ and T$_2$. Each of these modes can now be involved in the topological phase transition with the stationary zero mode, with a crossover point $g_A\approx 0.59, \gamma_{AB}\approx 0.17$. The modes continue to show all the required topological signatures in their correlation functions and stability excitation spectra. However, they all now display a larger mode volume, which is inherited from the profile of the zero mode in the linear case (cf.~Fig.~\ref{fig1}(b)). As a consequence, the output power of these modes (quantified by the intensities $I_A$ and $I_B$) has increased.

Compared to the situation in the SSH laser array in Fig.~\ref{fig3}, the modification of the gain imbalance examined
in Fig. \ref{figmirror2} now affects a much larger range of parameters, reaching up to $\gamma_{AB}\lesssim 2 g_B$. This can be attributed not only to the larger number of competing states, but also to the larger propensity of the zero mode to hybridize with such states in the central region, which on its own would constitute a topologically trivial system. In this regime we indeed encounter a very large number of additional solutions, which are all close to instability and therefore very sensitive to parameter changes, as demonstrated by the Bogoliubov-Floquet spectrum of the state marked P. Furthermore, an additional twisted mode appears close to the phase boundary of the zero mode, and indeed drives its instability along parts of this boundary (see the properties of the modes along the cross section at $\gamma_{AB}=0.2$).
In the remaining range of parameters, the system operates in analogous ways as before, with topological modes that display a larger output power when  compared to the SSH laser array with analogously reduced gain imbalance (Fig.~\ref{fig3}).

\section{\label{sec:results3}Robustness of operation conditions}

Typical bosonic systems are subject to fabrication imperfections and residual internal and external dynamics, which may or may not break the assumed symmetries.
For our laser arrays  \eqref{eq:model} with saturable gain \eqref{eq:potentials}, these deviations manifest themselves as linear static perturbations in the bare resonator frequencies $\omega_{s,n}$ and the couplings $\kappa_n$, $\kappa_n'$, and the symmetry-breaking nonlinearities quantified by the linewidth-enhancement factors $\alpha_s$.
We therefore consider the case of coupling disorder (with perturbations $\kappa_n=\bar\kappa (1+Wr_n)$, $\kappa_n'=\bar\kappa' (1+Wr_n')$) and onsite disorder (with perturbations $\omega_{A,n}=\omega_{AB}+Wr_n$,  $\omega_{B,n}=\omega_{AB}+Wr_n'$), where $r_n$, $r_n'$ are independent random numbers uniformly distributed in $[-1/2,1/2]$, and compare the effects with the case of a finite linewidth-enhancement factor $\alpha_A=\alpha_B$.

\subsection{Coupling disorder}
As a notable feature, the spectral and nonlinear dynamical symmetries of the considered laser arrays remain preserved if all perturbations are restricted to the couplings. This type of disorder does not affect the symmetry-protected spectral position of the defect mode in the linear model, and also preserves the classification of topological states in the nonlinear extension with saturable gain.

\begin{figure}[t]
\includegraphics[width=\columnwidth]{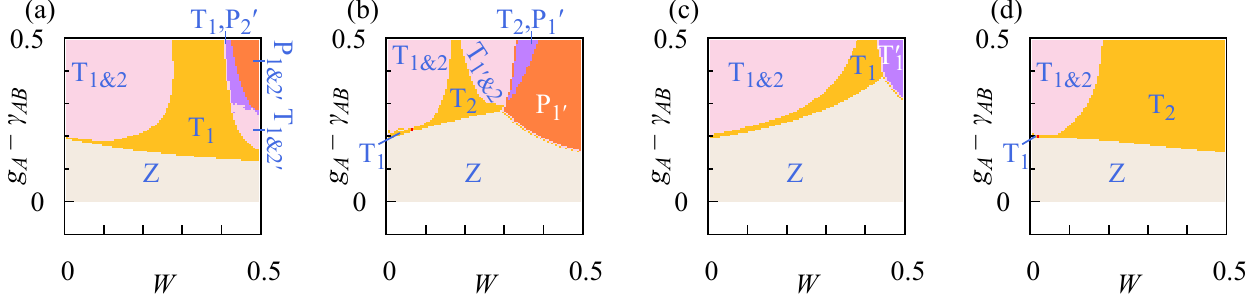}
\caption{\label{fig:tdisall} \textbf{Disorder-driven phase transitions} for the SSH laser array as in Figs.~\ref{fig2a}-\ref{fig2c}, but with fixed $\gamma_{AB}=0.1$ and variable strength $W$ of coupling disorder. Each panel corresponds to one randomly selected disorder configuration, with
perturbed couplings $\kappa_n=\bar\kappa (1+W r_n)$, $\kappa_n'=\bar\kappa' (1+W r_n')$ obtained from a fixed realizations of uniformly distributed random numbers  $r_n,r_n'\in [-1/2,1/2]$. Zero-mode lasing persists at all disorder strengths. Twisted states remain robust for weak to moderate disorder, while phase transitions to other operating regimes can appear when the disorder is very strong.
}
\end{figure}

As shown  for $W=0.1$ in Fig.~\ref{fig:tdisorder}  in the Appendix, small to moderate levels of coupling disorder have a practically negligible effect on the main operation regimes of the laser array. Such levels should be easily attainable in many applications, as they are well within the requirements to engineer any bandstructure effects in the first place.
Only at much larger strengths the fundamental effects of disorder become discernible.
As shown in Fig.~\ref{fig:tdisall}, this can result in disorder-strength-dependent phase transitions that modify the operation regimes in parts of parameter space, with the details generally depending on the disorder realization. Here, we have fixed the background losses to $\gamma_{AB}=0.1$, and instead vary the disorder strength for four fixed, randomly selected coupling profiles. In all cases, new operation regimes emerge only for very strong disorder $W\gtrsim 0.3-0.5$, so that the parameter space remains dominated by the zero mode and the two twisted states.

\begin{figure}[t]
\includegraphics[width=\columnwidth]{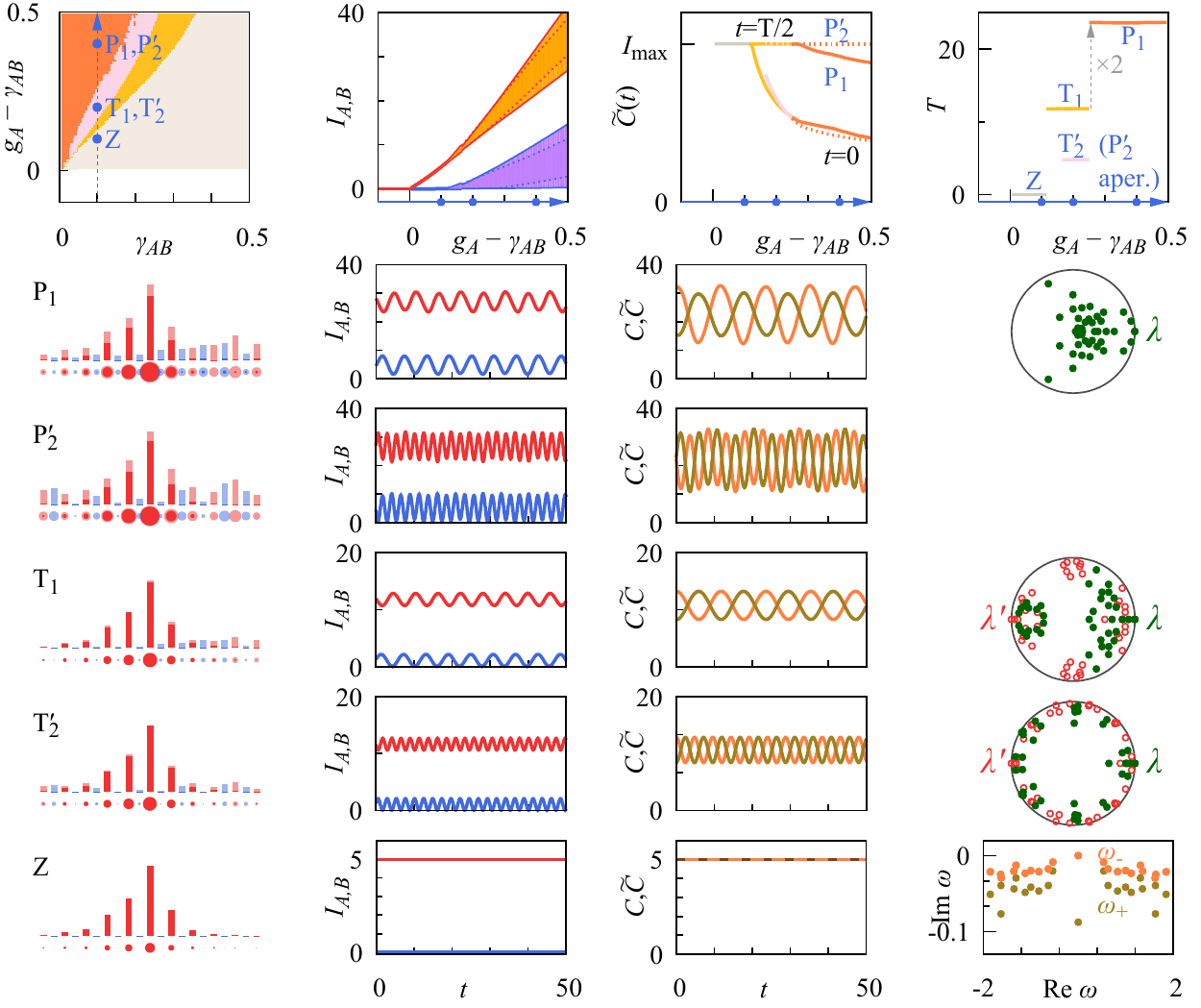}
\caption{\label{fig:t05disorder} \textbf{Effect of strong coupling disorder} for the SSH laser array as in Figs.~\ref{fig2a}-\ref{fig2c}, with the disorder configuration
of Fig.~\ref{fig:tdisall}(a) at $W=0.5$. For this realization the regime of zero-mode lasing is slightly reduced in favour of the power-oscillating twisted mode T$_1$, while the twisted state T$_2$ has been replaced by another twisted mode T$_2'$, which appears  in a disorder-strength-dependent phase transition. As gain is further increased, T$_1$ undergoes a period-doubling bifurcation to a symmetry-breaking pair of states P$_1$, while T$_2'$ is replaced by an aperiodic pair  P$_2'$ (for which the Floquet-Bogoliubov stability spectrum is not defined). All modes display visible distortions of their mode profile, and the symmetry-breaking pairs display noticeable amplitude on the B sublattice.
}
\end{figure}

\begin{figure}[t]
\includegraphics[width=\columnwidth]{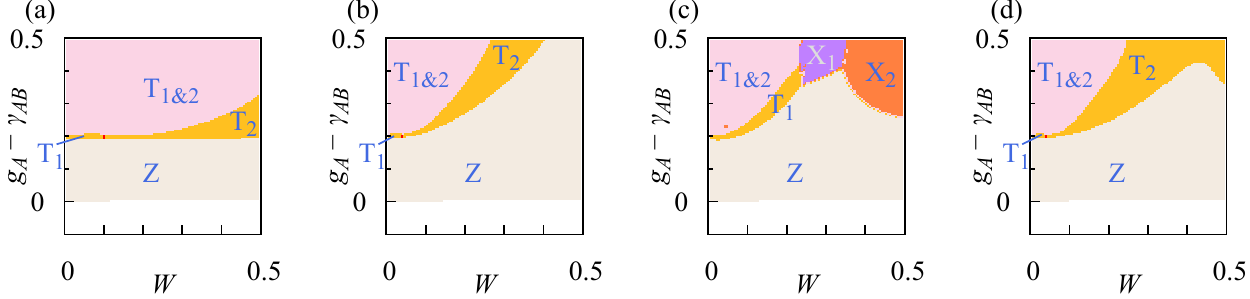}
\caption{\label{fig:wdisall} \textbf{Robustness against onsite disorder} in analogy to Fig.~\ref{fig:tdisall}, but for randomly selected disorder configurations with perturbed bare frequencies $\omega_{A,n}=\omega_{AB}+Wr_n$,  $\omega_{B,n}=\omega_{AB}+Wr_n'$,  $r_n,r_n'\in [-1/2,1/2]$. While this type of disorder breaks the symmetries, the states can typically be tracked to large values of disorder. The mode originating from the zero mode Z persists at all disorder strengths, and at weak to moderate disorder extends into regions of larger gain. This happens at the expense of the originally twisted modes, which in panel the configuration of (c) are replaced by new power-oscillating modes $X_1$, $X_2$ when the disorder becomes strong.
}
\end{figure}

\begin{figure}[t]
\includegraphics[width=\columnwidth]{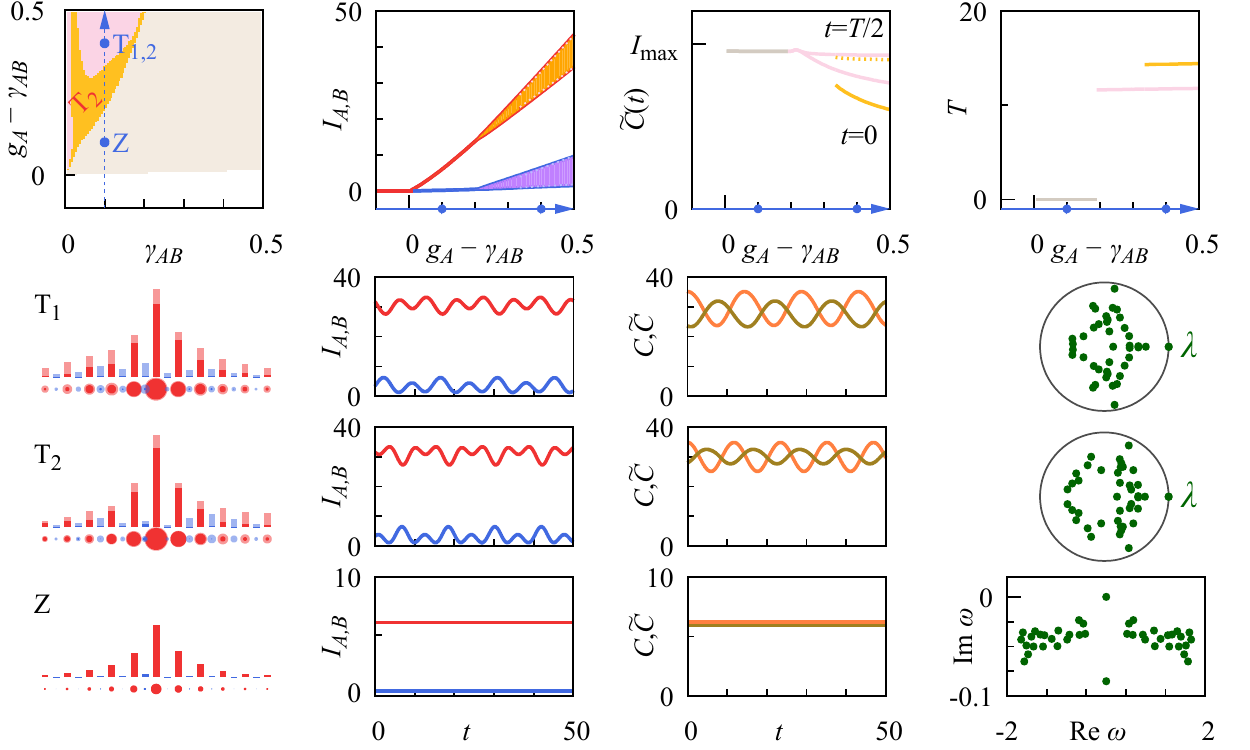}
\caption{\label{fig:w05disorder} \textbf{Effect of strong onsite disorder} in analogy to Fig.~\ref{fig:t05disorder}, for the disorder realization of Fig.~\ref{fig:wdisall}(a) at $W=0.5$. Even though the disorder breaks the symmetry, all states can be traced back to their disorder-free predecessors. The stationary lasing regime originating from the zero mode Z is barely affected. The mode originating from T$_1$ is pushed into a smaller part of parameter space, so that the instability phase transition now involves the modes originating from Z and T$_2$. The power-oscillations of the originally twisted states are modulated to clearly display the period $T$ of underlying amplitude oscillations. The mode profiles of all states are only slightly distorted.}
\end{figure}

\begin{figure}[t]
\includegraphics[width=\linewidth]{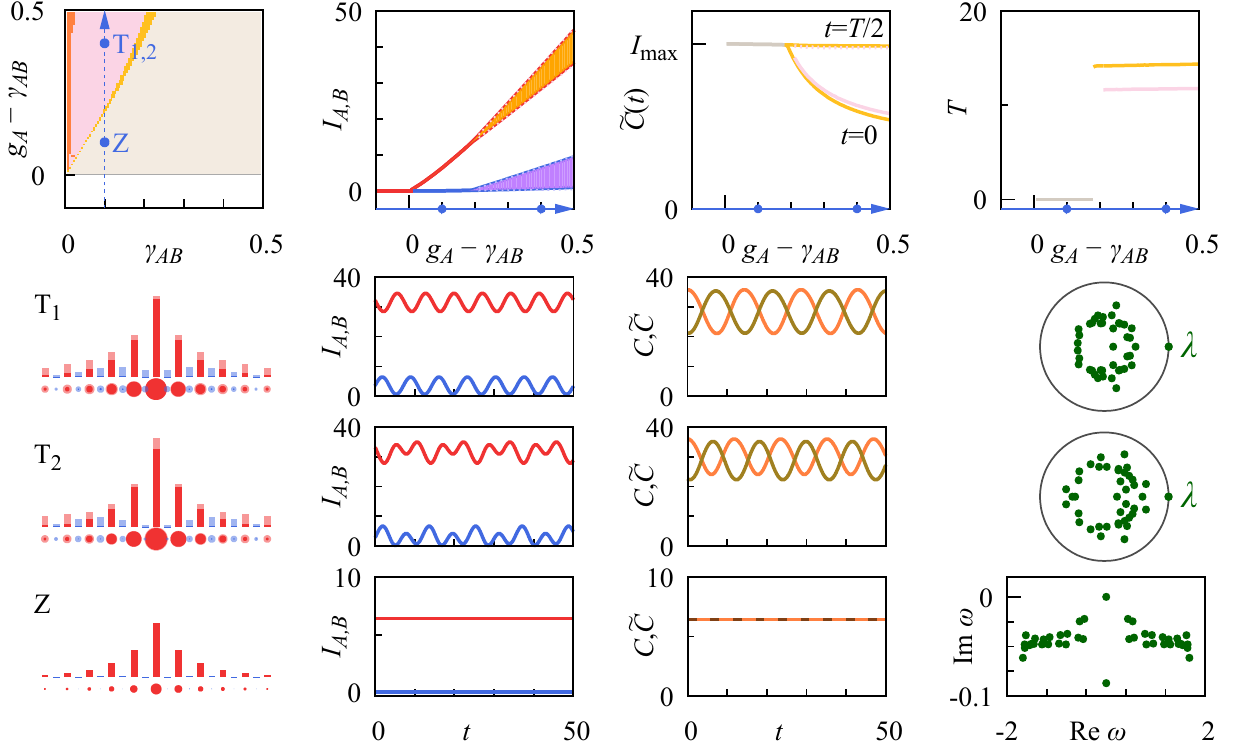}
\caption{\label{fig4}\textbf{Effect of nonlinear symmetry breaking} on the modes of the SSH laser array shown in Fig.~\ref{fig2a}-\ref{fig2c} (see also Fig.~\ref{fig:2summary}), obtained by setting the linewidth-enhancement factor to $\alpha_A=\alpha_B=0.5$. Most properties of the states are only slightly modified. The twisted correlation function $\tilde C(T/2)$ are slightly smaller than $I_{\rm max}$, while small independent modulations appear in the time-dependence of $C(t)$, $\tilde C(t)$. For the state originating from T$_2$, this results in noticeable modulations of the power oscillations, whose period is doubled.
There are also noticeable changes in the stability spectra (green), which can no longer be deconstructed as in the case of exact symmetry.}
\end{figure}

\begin{figure}[t]
\includegraphics[width=\linewidth]{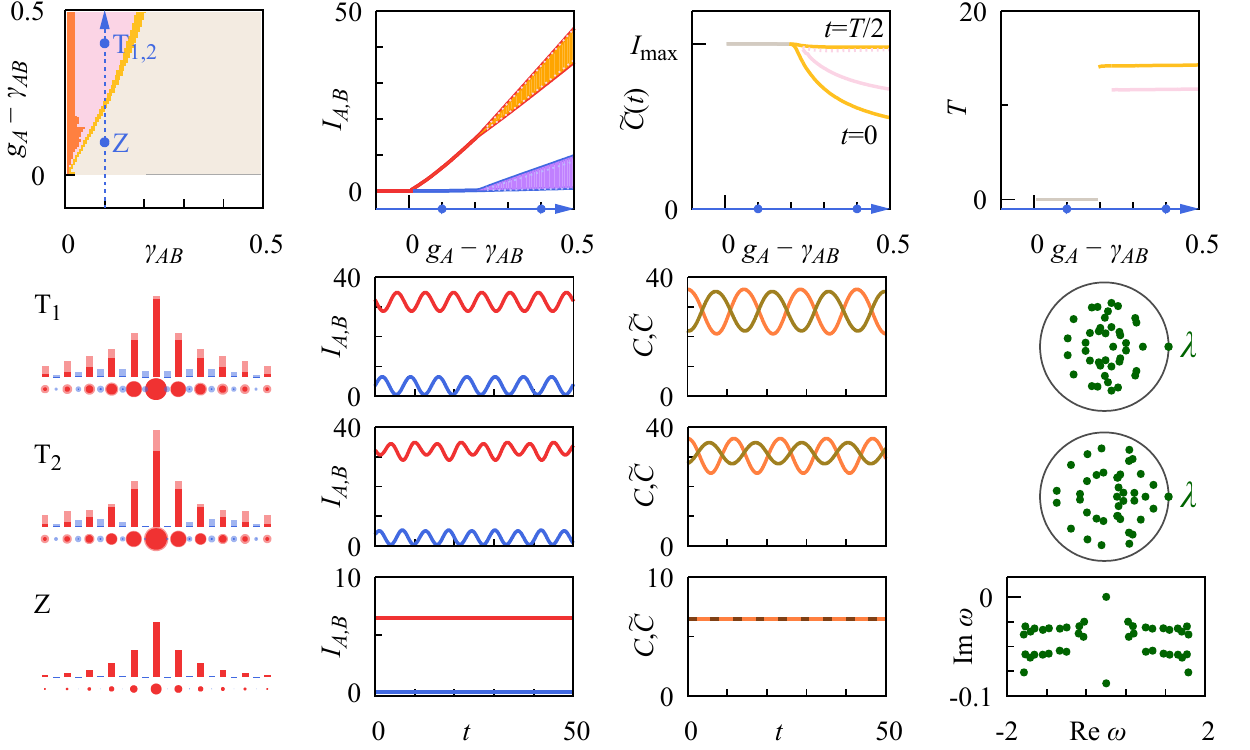}
\caption{\label{fig4a}\textbf{Effect of staggered nonlinear symmetry breaking} on the modes of the SSH laser array shown in Fig.~\ref{fig2a}-\ref{fig2c} (see also Fig.~\ref{fig:2summary}), obtained by setting the linewidth-enhancement factor to $\alpha_A=-\alpha_B=0.2$ so that the non-hermitian charge-conjugation symmetry is already broken in the linear regime. As in Fig. \ref{fig4}, most properties of the states are only slightly modified.}
\end{figure}

We further illustrate these emerging regimes in Fig.~\ref{fig:t05disorder}, which corresponds to the disorder configuration of
Fig.~\ref{fig:tdisall}(a) with   $W=0.5$.
As seen in this example, coupling disorder of this level can make all states visibly asymmetric and
push the power-oscillating twisted state T$_1$ into regions that previously supported the stationary zero mode Z, which however still dominates large parts of parameter space.
Even though here this twisted state has a period similar to T$_2$ in the clean case, it traces back to the state T$_1$ when the disorder strength is adiabatically reduced. The state labelled  T$_2'$, on the other hand, appears in a disorder-strength-dependent phase transition, and therefore cannot be traced back to any state in the clean system. Both twisted states become vulnerable to symmetry-breaking instabilities as one approaches conditions where the gain/loss ratio is large, $g_A\gg \gamma_{AB}$. In the given disorder realization, the twisted mode T$_1$ undergoes a period-doubling bifurcation into a pair of symmetry-breaking modes P$_1$, which goes along with a noticeable increase of weight on the B sublattice. The second twisted mode T$_2'$
also bifurcates into a pair of symmetry-breaking  modes, but these turn out to be aperiodic.

\subsection{Onsite disorder}

For onsite disorder, the strict classification of states by symmetry breaks down, and only the distinction between stationary states and power-oscillating states (as well as aperiodic and chaotic states) persists in a precise sense. However,
as shown for $W=0.1$ in Fig.~\ref{fig:wdisorder}  in the Appendix, the effects of small to moderate levels of onsite disorder are again barely noticeable,
just as in the case for coupling disorder. Furthermore, as shown in Fig.~\ref{fig:wdisall}, even
for relatively strong disorder the states can typically be traced back to their symmetry-respecting predecessors, which allows us to retain the previous labelling.
The disorder tends to expand the regime of stationary lasing originating from mode Z at the expense of the power-oscillating modes,
while only occasionally leading to transitions into new operation regimes.
Figure~\ref{fig:w05disorder} illustrates this resilience against strong disorder for the disorder configuration of Fig.~\ref{fig:wdisall}(a) with $W=0.5$.
For this disorder configuration the stationary lasing regime originating from mode Z is barely affected. Amongst the power oscillating states, the mode originating from T$_1$ is pushed into a smaller part of parameter space, so that the instability phase transition now involves the modes originating from Z and T$_2$.
The main visible consequence of broken symmetry is a modulation of the power-oscillations, which  now acquire the same period $T$ as the complex-amplitude oscillations, while the two correlation functions $C$ and $\tilde C$ exhibit different oscillation amplitudes.
Notably, the spatial intensity profiles of the states are still only slightly modified---indeed, they are affected more weakly than in the case of coupling disorder.

\subsection{Symmetry-breaking nonlinearities}

Similarly to the case of weak coupling and onsite disorder
we find that the lasing regimes are also highly resilient against realistic  symmetry-breaking nonlinearities, giving rise to practically negligible effects for $\alpha_A=\alpha_B=0.1$. As shown in Fig.~\ref{fig4}, even at much larger symmetry-breaking nonlinearities $\alpha_A=\alpha_B=0.5$ only small modifications are observed.
The effects of the nonlinearities are still small enough to preserve the division into stationary and power-oscillating states, even though the broken symmetry once more prevents the precise topological characterization of these states.
The symmetry-breaking terms again modulate the power oscillations, which is displayed more clearly for the mode originating from T$_2$. The Bogoliubov spectra show that the states remain highly stable as long as one stays away from the clearly defined phase transitions.
As shown in Fig.~\ref{fig4a}, this practical robustness also persists for a staggered arrangement with $\alpha_A=-\alpha_B=0.2$, which breaks the non-hermitian charge-conjugation symmetry already in the linear regime.

That this robustness persists both for symmetry-breaking onsite disorder and nonlinearities can be attributed to the spectral isolation of the defect mode in the linear model. This isolation suppresses any matrix elements of hybridization with extended modes in a perturbative treatment.
Note that in the linear case, this spectral isolation is increased by the favourable gain imbalance, as seen from the position of the complex resonance frequencies in the Fig.~\ref{fig1}. Furthermore, disorder can turn the extended modes into localized ones, thereby decreasing their mode volume.

\section{\label{sec:discussion}Discussion and conclusions}

The pursuit of topological effects in  photonic systems is motivated by the desire to achieve robust features in analogy to fermionic systems, which in the bosonic setting requires a dedicated effort to evoke the required symmetries.
The concept of a topological laser emerged from the realization that anomalous expectation values facilitate the selection of topological states by linear gain and loss.  Our investigation of topological laser arrays shows that these concepts seamlessly extend to the nonlinear setting, which accounts for the effects that stabilize active systems in their quasi-stationary operation regimes.
We uncovered large ranges in parameter space that favour topological operation conditions, of which we encountered two types---stationary lasing in self-symmetric zero modes, and lasing in twisted states displaying symmetry-protected power oscillations. The topological nature of these states can be ascertained by their characteristic spatial mode structure, and on a deeper level by distinctive properties of their correlation functions and linear excitation spectra. These features also uncover topological phase transitions in which zero modes and twisted states interchange their stability.
Encouragingly, the operation conditions can be tuned by changing the gain and loss distribution and the mode volume, while remaining remarkably robust under weak to moderate linear and nonlinear perturbations, even if these break the underlying symmetry.

These findings raise the prospect to explore the much simplified topological mode competition in a wide range of suitably patterned lasers with distributed gain and loss.  The laser arrays considered here and in the experiments \cite{St-Jean2017,zhao_topological_2018,PhysRevLett.120.113901} realize the required dynamical version of non-hermitian charge-conjugation symmetry by providing two sublattices, a setting that directly extends to two- and three-dimensional geometries, including systems with flat bands \cite{PhysRevLett.110.013903,2017arXiv170506895M}. Alternatively, one may also exploit orbital and  polarization degrees of freedom in suitably coupled multi-mode cavities, or design photonic crystals with an equivalent coupled-mode representation. By utilizing additional components that induce an imaginary vector potential (hence, directionally biased coupling), the mode competition in chains as studied here can be modified towards favouring a single extended states \cite{2018arXiv180100996L}, which further optimizes the mode volume.
All these systems promise to provide topological lasing modes with highly characteristic spatial and dynamical properties, which are stabilized at a working point that is spectrally well isolated from competing states in the system.

Looking beyond this symmetry class, it  will be worthwhile to explore the role of nonlinear distributed gain and loss in topological-insulator lasers \cite{Hararieaar4003,Bandreseaar4005}, where topological edge states align continuously along an edge band. This is a scenario which has been predicted to be more fragile against the carrier dynamics in the medium \cite{2018arXiv180406553L}, but is generally expected to benefit from non-hermitian effects, as has already been demonstrated for complex and directed coupling \cite{2018arXiv180205439L}. It would therefore be desirable to classify in general which nonlinearly extended dynamical symmetries can exist in these and other universality classes of topological systems, and whether this leads to novel operation regimes as described here for the case of non-hermitian charge-conjugation symmetry.

We gratefully acknowledge enlightening discussions with Ramy El-Ganainy, Emiliano Cancellieri, and Takahisa Harayama, as well as
support by EPSRC via Programme Grant No. EP/N031776/1 and Grant No. EP/P010180/1.

\appendix

\section{Summary of results for reference}

\begin{figure}[t]
\includegraphics[width=\columnwidth]{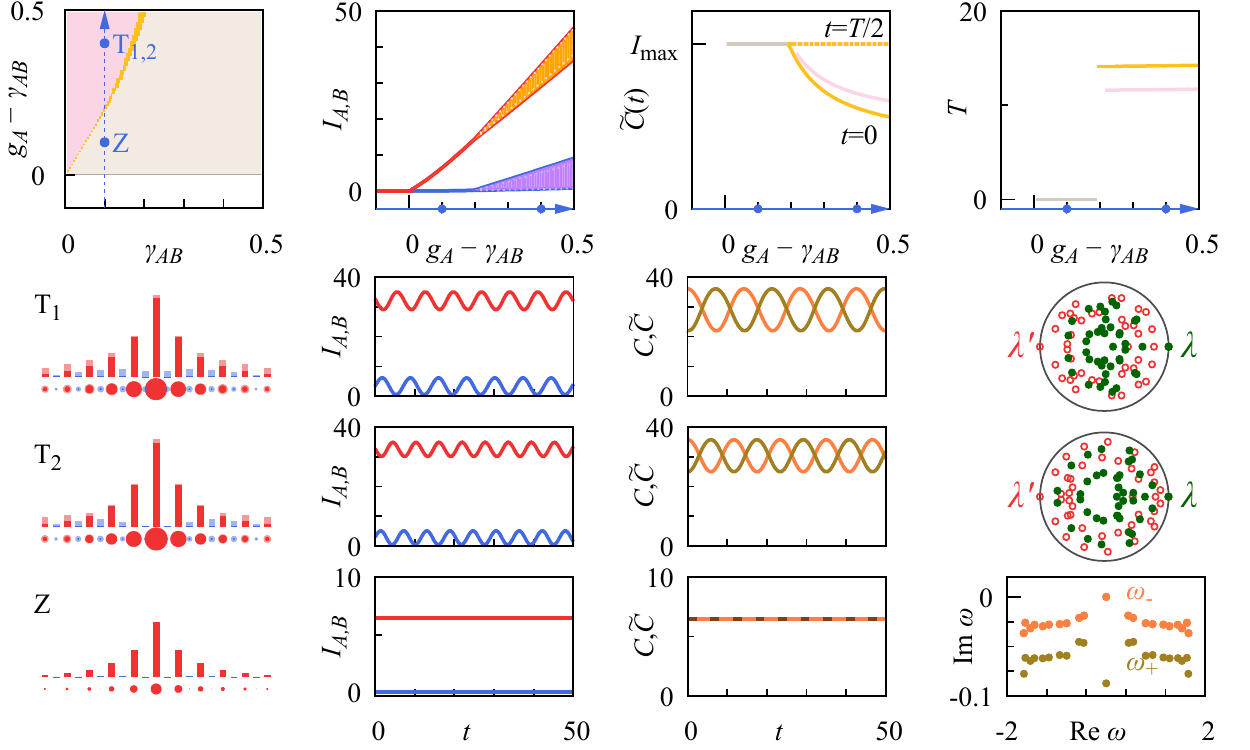}
\caption{\label{fig:2summary} \textbf{Overview of results for the ideal SSH laser}, summarizing Figs.~\ref{fig2a},
\ref{fig2b}, and \ref{fig2c}, for reference and comparison with the condensed figures for other operation conditions in the main text.
}
\end{figure}

For reference, Fig.~\ref{fig:2summary} summarizes the results of Figs.~\ref{fig2a},
\ref{fig2b}, and \ref{fig2c} for ideal topological lasing in the same format as adopted in the figures for the other operation conditions covered in this work.

\section{Resilience against weak perturbations}
\begin{figure}[t]
\includegraphics[width=\columnwidth]{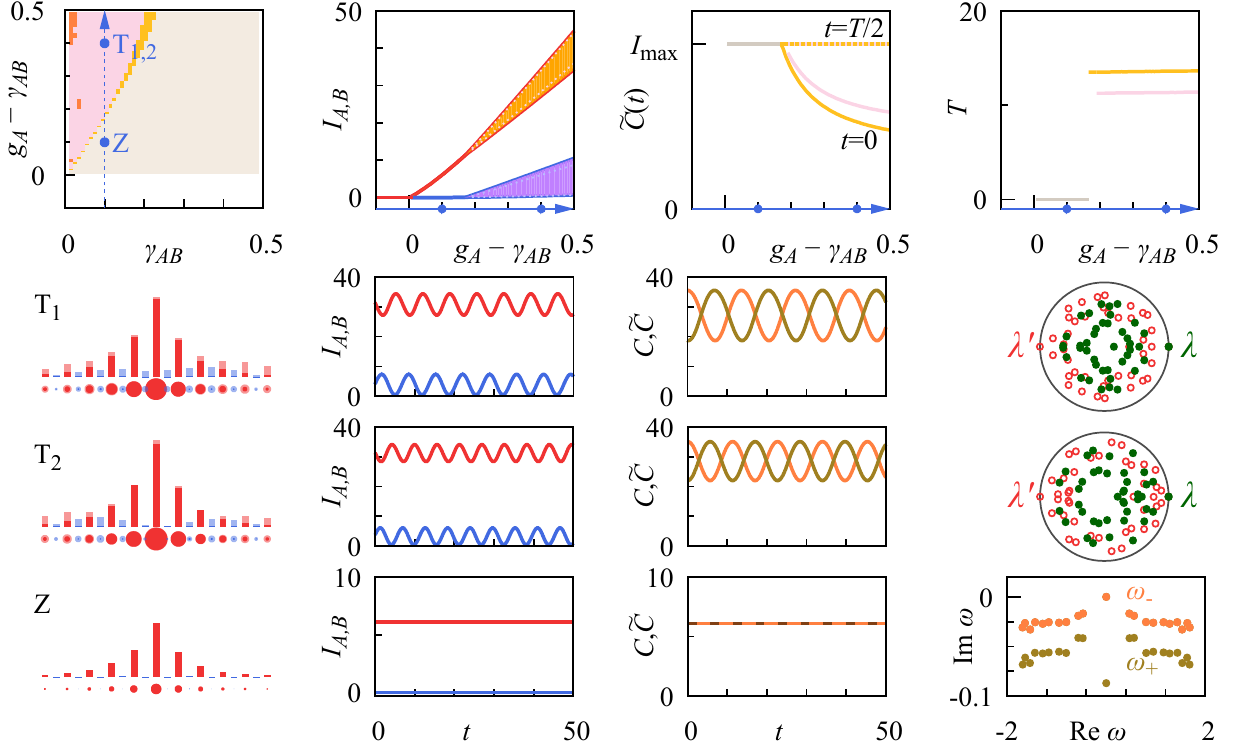}
\caption{\label{fig:tdisorder} \textbf{Effect of weak coupling disorder} on the SSH laser array, obtained for a representative disorder realization with $W=0.1$ (see Fig.~\ref{fig:t05disorder} for the analogous results with $W=0.5$). The results are virtually identical to those in the ideal system (summarized in Fig.~\ref{fig:2summary}).}
\end{figure}

\begin{figure}[t]
\includegraphics[width=\columnwidth]{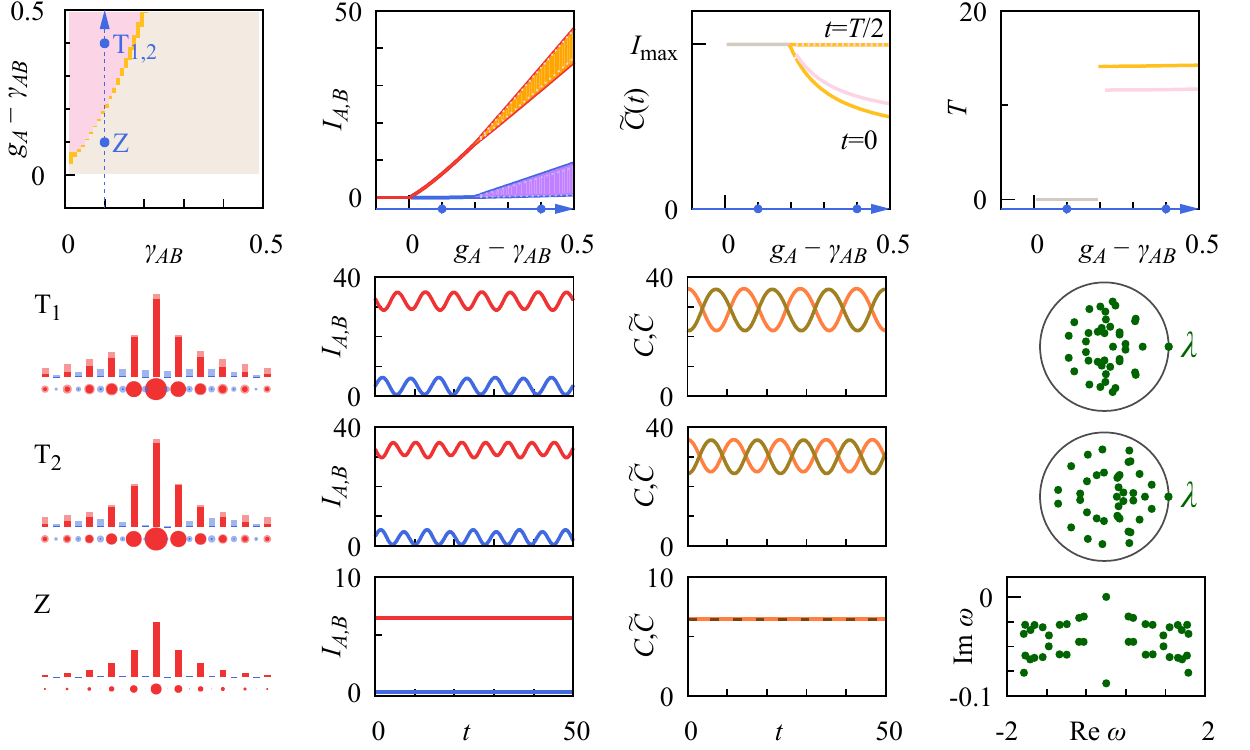}
\caption{\label{fig:wdisorder} \textbf{Effect of weak onsite disorder} on the SSH laser array for a representative disorder realization with $W=0.1$
(see Fig.~\ref{fig:w05disorder} for the analogous results with $W=0.5$). As in the case of coupling disorder (Fig.~\ref{fig:tdisorder}), the results are  virtually identical to those in the ideal system (summarized in Fig.~\ref{fig:2summary}).}
\end{figure}

As mentioned in the Sect.~\ref{sec:results3}, weak to moderate amounts of disorder have a negligible effect on the operations regimes. This is illustrated for coupling disorder in Fig.~\ref{fig:tdisorder} and for onsite disorder in  Fig.~\ref{fig:wdisorder}, where in both cases $W=0.1$.

\section{Bogoliubov theory}

\subsection{Preparations}

In matrix form, the nonlinear evolution equations \eqref{eq:model} can be written as
\begin{align}\label{eq:modelmatrix}
i\frac{d}{dt} \Psi(t)&=H\Psi(t)+V[\Psi(t)]\Psi(t),\quad\Psi(t)=\left(\begin{array}{c}\mathbf{A}(t)\\ \mathbf{B}(t)\end{array}\right),\\
H&=\left(\begin{array}{cc}\omega_A   & K \\ K^T & \omega_B\end{array}\right), \quad
V[\Psi]=\left(\begin{array}{cc}V_A & 0 \\ 0 & V_B \end{array}\right),
\end{align}
where \begin{equation}
K_{nm}=\delta_{nm}\kappa_n+\delta_{n,m+1}\kappa'_n
\end{equation}
 represents the couplings, while the resonance frequencies and nonlinear potentials  (corresponding to Eq.~\eqref{eq:potentials}) have been promoted to diagonal matrices,
\begin{align}
\omega_{A,nm}=\delta_{nm}\omega_{A,n},\quad \omega_{B,nm}=\delta_{nm}\omega_{B,n},\\
V_{A,nm}=\delta_{nm}V_{A,n},\quad V_{B,nm}=\delta_{nm}V_{B,n}.
\end{align}
Stationary states $\Psi(t)=\exp(-i\Omega_n t)\Psi^{(n)}$ with real frequency $\Omega_n$ are determined as self-consistent solutions of the equation
\begin{equation}
\label{eq:stat}
\Omega_n\Psi^{(n)}=(H+V[\Psi^{(n)}])\Psi^{(n)},
\end{equation}
while general periodic states of period $T$  fulfill
\begin{equation}
\label{eq:periodic}
\Psi(T)=\exp(-i\varphi)\Psi(0)
\end{equation}
with a real phase $\varphi$.

As in the main text, we set the reference frequency $\omega_{AB}=0$ [a finite value can always be reinstated by multiplying any solution by $\exp(-i\omega_{AB}t)$].
The property
\begin{equation}
(H+V)^*|_{\omega_{s,n},\alpha_s}=-\sigma_z(H+V)\sigma_z|_{-\omega_{s,n},-\alpha_s}
\end{equation}
with $\sigma_z=\left(\begin{array}{cc}1 & 0 \\ 0 & -1 \end{array}\right)$
 then results in the mapping of solutions
\begin{equation}
\label{eq:tildepsi2}
\tilde \Psi(t)|_{-\omega_{s,n},-\alpha_s}= \sigma_z\Psi(t)|_{\omega_{s,n},\alpha_s},
\end{equation}
cf.~Eq.~\eqref{eq:tildepsi}.
For $\alpha_s=0$, $\omega_{s,n}=0$, this becomes a statement for solutions within a fixed set of parameters.

In the purely linear case with effective potentials $V_A=i(g_A-\gamma_A)\equiv iG_A$, $V_B=i(g_B-\gamma_B)\equiv iG_B$, we encounter the conventional non-hermitian charge-conjugation symmetry  $\sigma_zH_0\sigma_z=-H_0^*$ for the linear Hamiltonian $H_0=H+V$ \cite{schomerus_topologically_2013,poli_selective_2015}. We can then exploit that $K$ is an $(N+1)\times N$-dimensional matrix (as there is one more A site than B sites) to determine one zero mode with $K^T\textbf{A}^{(Z)}=0$, $\textbf{B}^{(Z)}=0$ \cite{PhysRevB.34.5208}. This mode  obeys $H_0\Psi^{(Z)}=iG_A\Psi^{(Z)}$, which above threshold ($G_A>0$) describes an exponentially increasing state, signifying the lack of feedback in the linear theory. In this linear case, the extended states still occur in pairs $\Psi_n$, $\tilde \Psi_n$ with generally complex $\Omega_n=-\tilde \Omega_n^*$, unless $\mathrm{Re}\,\Omega_n=0$, which describes additional self-symmetric states that can occur via a spectral phase transition \cite{Pikulin2012a,PhysRevB.87.235421,malzard_topologically_2015,sanjose2016}.  Using optical reciprocity, $H_0=H_0^T$, the symmetry-breaking states are constrained by the condition
\begin{align}
0&=\Psi_n^\dagger (\sigma_z H_0+H_0^\dagger \sigma_z)\Psi_n=(\Omega_n+\Omega_n^*)\Psi_n^\dagger\sigma_z\Psi_n\nonumber\\
&=(\Omega_n+\Omega_n^*)(|\mathbf{A}|^2-|\mathbf{B}|^2),
\end{align}
hence  $|\mathbf{A}|=|\mathbf{B}|$.
Furthermore, from
\begin{align}
&i\Psi_n^\dagger (G_A+G_B)\Psi_n
=
i\Psi_n^\dagger [G_A+G_B +\sigma_z(G_A-G_B)]\Psi_n
\nonumber
\\
&
=
\Psi_n^\dagger (H_0-H_0^\dagger)\Psi_n
=(\Omega_n-\Omega_n^*)\Psi_n^\dagger\Psi_n,
\end{align}
we find that they all have the same life time, according to $\mathrm{Im}\,\Omega_n=(G_A+G_B)/2\equiv\bar G$.
This confirms the statements in Sect.~\ref{sec:linselection} and Fig.~\ref{fig1}.

In the nonlinear case, the relation between solutions at fixed parameters applies to stationary zero modes
\begin{equation}
\label{eq:z}
\Psi^{(Z)}=\tilde \Psi^{(Z)},
\end{equation}
which now must be stabilized at an exactly vanishing frequency $\Omega_Z=0$ [see Eq.~\eqref{eq:stat}], and twisted modes
\begin{equation}
\Psi^{(T)}(T/2)=\tilde \Psi^{(T)}(0).
\label{eq:t}
\end{equation}
For both  cases, these definitions exploit the U(1) gauge freedom to multiply any solution by an overall phase factor $\exp(i\chi)$. E.g., if a zero mode fulfills
$\Psi^{(Z)\prime}=\exp(-2i\chi)\tilde \Psi^{(Z)\prime}$ then $\Psi^{(Z)}=\pm\exp(i\chi)\Psi^{(Z)\prime}$ fulfills Eq.~\eqref{eq:z}, and the same redefinition applies for a twisted mode $\Psi^{(T)\prime}(T/2)=\exp(-2i\chi)\tilde \Psi^{(T)\prime}(0)$.
Irrespective of these redefinitions, zero modes always display a rigid phase difference of $\pm\pi/2$ between the amplitudes on the A and the B sublattice, while
twisted modes always fulfill $\Psi(T)=\Psi(0)$, i.e. they are periodic modes \eqref{eq:periodic} with guaranteed $\varphi=0$.

\subsection{Stability analysis}
Given a reference solution $\Psi(t)$ of the nonlinear wave equation \eqref{eq:modelmatrix}, we can analyse its stability by adding a small perturbation
\begin{equation}
\delta\Psi(t)=u(t) + v^*(t), \quad u=\left(\begin{array}{c}u_A(t)\\u_B(t)\end{array}\right),\quad  v=\left(\begin{array}{c}v_A(t)\\v_B(t)\end{array}\right),
\end{equation}
and linearizing in $u$ and $v$.
This yields the Bogoliubov equation
 \begin{equation}
i\frac{d}{dt} \psi(t)
=\mathcal{H}[\Psi(t)]\psi(t),\quad \psi(t)=\left(\begin{array}{c}u_A(t)\\u_B(t)\\ v_A(t)\\v_B(t)\end{array}\right),
\label{eq:timedepB}
\end{equation}
with the Bogoliubov Hamiltonian
\begin{align}
\label{eq:bdg}
\mathcal{H}[\Psi]&=
\left(\begin{array}{cc}
H+\Gamma& \Delta\\
-\Delta^* & -H^*-\Gamma^*
\end{array}\right),
\\
\Gamma&=
\left(\begin{array}{cc}
\Gamma_A& 0\\
0 & \Gamma_B
\end{array}\right),
\quad
\Delta=
\left(\begin{array}{cc}
\Delta_A& 0\\
0 & \Delta_B
\end{array}\right),
\end{align}
where
\begin{align}
\Gamma_{A,nm}&=\delta_{nm}(i+\alpha_A)
\left(\frac{g_A}{(1+S_A|A_n|^2)^2}-\gamma_A\right),
\\
\Gamma_{B,nm}&=\delta_{nm}(i+\alpha_B)
\left(\frac{g_A}{(1+S_B|B_n|^2)^2}-\gamma_B\right),
\\
\Delta_{A,nm}&=-\delta_{nm}
(i+\alpha_A)\frac{S_Ag_AA_n^2}{(1+S_A|A_n|^2)^2},
\\
\Delta_{B,nm}&=-\delta_{nm}
(i+\alpha_B)\frac{S_Bg_BB_n^2}{(1+S_B|B_n|^2)^2}.
\end{align}

For a stationary state fulfilling Eq.~\eqref{eq:stat}, we seek solutions of the form $u_s=\exp(-i\Omega_nt-\omega_m t)u_{s}^{(m)}$,
$v_s=\exp(i\Omega_nt-\omega_m t)v_{s}^{(m)}$ ($s=A,B)$, which follow from the eigenvalue equation
\begin{equation}\label{eq:statB}
\omega_m \psi^{(m)} =\left(\mathcal{H}[\Psi^{(n)}]-  \Omega_n \Sigma_z\right) \psi^{(m)}
\end{equation}
where here and in the following we use the Pauli-like matrices
\begin{equation}
\Sigma_x=\left(\begin{array}{cccc}0&0&1&0\\0&0&0&1\\1&0&0&0\\0&1&0&0\end{array}\right), \quad
\Sigma_z=
\left(\begin{array}{cccc}
1 & 0 & 0 & 0 \\
0 & 1 & 0 & 0 \\
0 & 0 & -1 & 0 \\
0 & 0 & 0 & -1 \\
\end{array}\right).
\end{equation}
For a periodic state \eqref{eq:periodic}, we first integrate the Bogoliubov equation over a period, so that
$\psi(T)=U(T)\psi(0)$, and then introduce the Bogoliubov-Floquet operator
\begin{equation}
F=\exp(i\Sigma_z\varphi)U(T),
\end{equation}
whose eigenvalues are denoted as $\lambda_m=\exp(-i\omega_m T)$. Here, the shift by the phase factor $\varphi$ plays a similar role as the appearance of $\Omega_n$ in Eq.~\eqref{eq:statB}. In both cases, a solution is stable if all eigenvalues fulfill $\mathrm{Im}\,\omega_m\leq 0$, so that the associated perturbations do not grow over time.

In general, the Bogoliubov Hamiltonian displays the symmetry
\begin{equation}
\label{eq:gensym}
(\mathcal{H}[\Psi])^*=-\Sigma_x\mathcal{H}[\Psi]\Sigma_x.
\end{equation}
In the stationary case, this yields a spectrum $\omega_m$ that is symmetric under reflection about the imaginary axis,
yielding pairs of eigenvalues
$\omega_m$, $\tilde\omega_m=-\omega_m^*$ and individual purely imaginary  eigenvalues
$\omega_m= -\omega_m^*$. This includes a U(1) Goldstone mode
\begin{equation}
\label{eq:u1}
\psi^{(0)}=\left(\begin{array}{c}\Psi^{(n)}\\-\Psi^{(n)*}\end{array}\right),\quad  \omega_0=0,
\end{equation}
which accounts for the free choice of the overall phase factor of a stationary solution.
Analogously, the Bogoliubov-Floquet spectrum contains complex-conjugate pairs of eigenvalues  $\lambda_m$, $\tilde \lambda_m=\lambda_m^*$ and individual  real  eigenvalues $\lambda_m=\lambda_m^*$. This again includes
a U(1) Goldstone mode
\begin{equation}
\label{eq:u1a}
\psi^{(0)}=\left(\begin{array}{c}\Psi(t)\\-\Psi^*(t)\end{array}\right),\quad  \lambda_0=1
\end{equation}
 reflecting the free choice of the overall phase of any solution, and now also a time-translation Goldstone mode
\begin{equation}
\label{eq:u2}
\psi^{(t)}=\left(\begin{array}{c}d\Psi/dt\\d\Psi^*/dt\end{array}\right),\quad  \lambda_t=1
\end{equation}
that reflects the freedom to displace any solution $\Psi(t)$ in time.

\subsection{Topological modes}

To account for the possible symmetries of the nonlinear evolution equation \eqref{eq:modelmatrix} we adapt the general considerations of \cite{2017arXiv170506895M}.
The mapping of solutions \eqref{eq:tildepsi2} amounts to the property
\begin{align}\label{eq:bgsym}
\left.(\mathcal{H}[\Psi])^*\right|_{\omega_{s,n},\alpha_s}&=
-\left.
\mathcal{Z}\mathcal{H}[\tilde\Psi]\mathcal{Z}
\right|_{-\omega_{s,n},-\alpha_s},\\ \mathcal{Z}&=\left(\begin{array}{cc}
\sigma_z& 0\\
0 & \sigma_z
\end{array}\right).
\end{align}
Along with Eq.~\eqref{eq:gensym}, this property dictates that the Bogoliubov excitation spectra of the two mapped solutions $\Psi$, $\tilde\Psi$ are identical. For $\alpha_s=0$, $\omega_{s,n}=0$, we can use this to further deconstruct the excitation spectra of topological modes.
For zero modes \eqref{eq:z}, we can distinguish symmetry-preserving excitations $v_A=u_A$, $v_B=-u_B$,
fulfilling
\begin{equation}
\omega_{+,m} u^{(+,m)} = (H+2\Gamma-V) u^{(+,m)},
\end{equation}
from symmetry-breaking excitations $v_A=-u_A$, $v_B=u_B$, fulfilling
\begin{equation}
\omega_{-,m} u^{(-,m)} = (H+V) u^{(-,m)},
\end{equation}
where the latter includes the mode \eqref{eq:u1}, now expressed as $u^{(-,0)}=\Psi^{(n)}$, $\omega_{-,0}=0$.

For twisted modes \eqref{eq:t}, we can factorize the Bogoliubov-Floquet propagator
\begin{align}
F&=\mathcal{Z}U^*(T/2)\mathcal{Z}U(T/2)
\nonumber \\
&=\mathcal{Z}\Sigma_xU(T/2)\Sigma_x\mathcal{Z}U(T/2)
\nonumber \\
&={F'}^2,\quad
 \\
F'&=\mathcal{Z}\Sigma_xU(T/2),
\end{align}
which defines the twisted half-step propagator $F'$. Its eigenvalues $\lambda_m'$ determine the stability spectrum  as $\lambda_m=(\lambda_m')^2$. The U(1) Goldstone mode \eqref{eq:u1a} fulfills
$\psi^{(0)}(T/2)=-\mathcal{Z}\Sigma_x\psi^{(0)}(0)$, so that the associated eigenvalue $\lambda_0'=-1$, while the time-translation mode \eqref{eq:u2}  fulfills $\psi^{(t)}(T/2)=\mathcal{Z}\Sigma_x\psi_T(0)$, so that $\lambda_t'=1$.

\subsection{A brief note on time evolution}
The Bogoliubov Hamiltonian \eqref{eq:bdg} also naturally appears in an efficient numerical integration scheme of the nonlinear wave equation \eqref{eq:modelmatrix}. For this we first introduce the wave equation in the doubled space,
\begin{align}
i\dot \Phi&=\mathcal{H}_0\Phi,\quad
\Phi=\left(\begin{array}{c}\Psi\\ \Psi^*\end{array}\right),\\
 \mathcal{H}_0&=\left(\begin{array}{cc}H +V& 0 \\ 0 & -H^*-V^*\end{array}\right).
\end{align}
Using the mid-point predictor
\begin{equation}
\Phi(t+dt)\approx(1-idt H-idt V[\Phi(t+dt/2)])\Phi(t)
\end{equation}
and linearizing in the exact same way as in the stability analysis, we then obtain
\begin{equation}
\Phi(t+dt)\approx(1+i \mathcal{H}dt/2 )^{-1}[1-i (2\mathcal{H}_0-\mathcal{H})dt/2 ]\Phi(t),
\end{equation}
which amounts to a second-order integrator akin to the Crank-Nicolson scheme.

%\bibliography{schomerus}
%merlin.mbs apsrev4-1.bst 2010-07-25 4.21a (PWD, AO, DPC) hacked
%Control: key (0)
%Control: author (0) dotless jnrlst
%Control: editor formatted (1) identically to author
%Control: production of article title (0) allowed
%Control: page (1) range
%Control: year (0) verbatim
%Control: production of eprint (0) enabled
%

\end{document}